# Two-dimensional perovskites with maximum symmetry enable exciton diffusion length exceeding 2 micrometers

Jin Hou[1,#], Jared Fletcher[2,#], Siedah J. Hall[3,4], Hao Zhang[5,6], Marios Zacharias[7], George Volonakis[8], Claire Welton[9], Faiz Mandani[5], Isaac Metcalf[1], Shuo Sun[10], Bo Zhang[10], Yinsheng Guo[10], G. N. Manjunatha Reddy[9], Claudine Katan[8], Jacky Even[7], Matthew Y. Sfeir[3,4], Mercouri G. Kanatzidis[2*] and Aditya D. Mohite[1,5*]

[1]Department of Materials Science and NanoEngineering, Rice University, Houston, Texas 77005, USA.

[2]Department of Chemistry and Department of Materials Science and Engineering, Northwestern University, Evanston, Illinois 60208, USA.

[3]Photonics Initiative, Advanced Science Research Center, City University of New York, New York, 10031, USA

[4]Department of Physics, The Graduate Center, City University of New York, New York, 10016, USA

[5]Department of Chemical and Biomolecular Engineering, Rice University, Houston, Texas 77005, USA.

[6]Applied Physics Graduate Program, Smalley-Curl Institute, Rice University, Houston, TX, 77005, USA.

[7]Univ Rennes, INSA Rennes, CNRS, Institut FOTON - UMR 6082, 35708 Rennes, France.

[8]Univ Rennes, ENSCR, INSA Rennes, CNRS, ISCR (Institut des Sciences Chimiques de Rennes) - UMR 6226, F-35000 Rennes, France.

[9]University of Lille, CNRS, Centrale Lille Institute, Univ. Artois, UMR 8181−UCCS−Unité de Catalyse et Chimie du Solide, F-59000, Lille, France.

[10]Department of Chemistry, University of Nebraska-Lincoln, Lincoln, NE 68588

#Contributed equally
*Correspondence: m-kanatzidis@northwestern.edu and adm4@rice.edu

**Realizing semiconductors with high symmetry of their crystallographic structures has been a virtue of inorganic materials and has resulted in novel physical behaviors. In contrast, hybrid (organic and inorganic) crystals such as two-dimensional metal halide perovskites exhibit much lower crystal symmetry due to in-plane or out-of-plane octahedral distortions. Despite their amazing ability for photoinduced light emission at**

**room temperature, the Achilles' heel of this attractive class of 2D materials for optoelectronics remains the poor control and lack of performance for charge carrier transport. Inspired by the tremendous charge carrier properties of the 3D cubic perovskite phase (α) of FAPbI$_3$ and combining the use of the appropriate cage cation, the spacer molecule and the temperature and rate of crystallization, we report a new series of FA-based layered two-dimensional perovskites that exhibits the highest theoretically predicted symmetry with a tetragonal *P4/mmm* space group, resulting in no octahedral distortion in both in-plane and out-of-plane directions. These 2D perovskites present the shortest interlayer distances (4 Å), which results in systematically lower bandgaps (1.7 to 1.8 eV). Finally, the absence of octahedral distortions, results in an exciton diffusion length of 2.5 μm, and a diffusivity of 4.4 cm$^2$s$^{-1}$, both of which are an order of magnitude larger compared to previously reported 2D perovskites and on par with monolayer transition metal dichalcogenides.**

Symmetry is a foundationally important concept, which influences our understanding of the fundamental laws of the universe. For example, symmetries are directly linked to fundamental interactions, conservation laws and govern selection rules. Symmetry also plays an important role in chemistry and biology, being responsible, for example, for the reactivity of a specific molecular enantiomer. In materials science, symmetry breaking is key to our understanding of phase transitions, which are identified by its lattice symmetry and are directly linked to the discovery and development of many physical behaviors, such as altermagnetism,[1] high harmonic generation,[2] and topological materials.[3] The high crystal symmetry of semiconductors such as Silicon (diamond cubic) and GaAs (zinc blende) drastically influences their electronic, optical, and thermal properties. In Silicon, symmetry leads to an indirect bandgap, reducing optical absorption but enabling high carrier mobility and thermal conductivity due to reduced phonon scattering. In GaAs, symmetry allows a direct bandgap, making it ideal for optoelectronics with efficient light emission and higher electron mobility. These properties have enabled Silicon for its use in microelectronics, and GaAs in photonic and high-speed applications.

However, in contrast, materials with a soft lattice, such as organic-inorganic (hybrid) halide perovskites (perovskites hereafter), the crystal symmetry is often lowered due to octahedra tilting, which results from the process of relieving lattice strain to prevent bond breaking. For example, among the most studied three-dimensional (3D) perovskites, FAPbI$_3$ (FA=formamidinium) and MAPbI$_3$ (MA = methylammonium) exhibit a non-perovskite hexagonal phase and a tetragonal perovskite phase at room temperature (RT), respectively. Both transform into a higher symmetric perovskite phase only at elevated temperature. In the non-perovskite phase of FAPbI3, charge transport properties are poor due to the high density of edge-sharing octahedra (also known as delta phase or Yellow phase).[4] In contrast, the cubic perovskite phase (α) of FAPbI3, which features corner-sharing octahedra, exhibits electron or hole diffusion lengths ranging from 6.6 µm to 600 µm[5,6]. For comparison, the lower-symmetry tetragonal phase of MAPbI3 has electron or hole diffusion lengths between approximately 10 and 170 µm.[7,8] While these are about an order of magnitude smaller than Silicon[9], they are exceptional for solution-processed thin films and have led to highly efficient optoelectronic devices. However, 3D perovskites suffer from stability issues, specifically under humidity, heat, and light, which can induce ion transport within the film and deteriorate performance.[10,11]

Two-dimensional (2D) perovskites, on the other hand, have shown to offer a viable solution to the stability challenge due to the intercalation of hydrophobic organic cations, forming an organic-inorganic superlattice structure. They can be described with a general formula of $(A')_m(A)_{n-1}M_nX_{3n+1}$ (where A′ is a large organic cation, A is a small organic cation (e.g., MA and FA), M is a divalent metal (e. g. Pb), X is a halide (e.g. I), and n determines the thickness of the $(A)_{n-1}M_nX_{3n+1}$ perovskite layer). The large size of the organic A' site cations induces excess strain, which often results in octahedron distortions (in-plane or out-of-plane) at the boundary of the inorganic-organic layers[12,13], making it challenging to achieve high symmetry in 2D perovskites. To date, there have been no reports of 2D halide perovskite with P4/mmm tetragonal phase (the theoretical highest symmetry for 2D layered structures) at room temperature and pressure and limited reports at temperatures >100 ˚C.[14,15] However, the carrier mobility (for 3D perovskites) or exciton diffusivity (for 2D perovskites) generally decreases as temperature increases (in the regime above room temperature), due to enhanced exciton-phonon interactions and other scattering mechanisms.[16–18] In addition, higher temperatures accentuate other undesired effects such as ion migration, further limiting device performance.[19] Moreover in 2D perovskites, instead of free carriers (electrons or holes), excitons are the dominant excitation species due to their large exciton binding energies arising from quantum and dielectric confinement[20] thus making it important to understand exciton transport. Specifically, from the context of minority carrier devices such as photovoltaics and photodetectors, long exciton diffusion length is required for excitons to reach interfaces to be dissociated and separated into charge carriers. However, the exciton transport in 2D perovskites has been shown to be greatly influenced by the structural symmetry and degree of lattice distortions. It has been experimentally validated that large structural distortions of the 2D perovskite lattice resulting from octahedral tilting (described by deviations of Pb-I-Pb angles from 180°) are correlated with short exciton diffusion lengths.[21] These distortions reduce the orbital overlapping, which increases the effective mass.[22,23] Seitz et al. showed that the exciton diffusion length of $(PEA)_2PbI_4$ (236 nm) is much higher than in $(BA)_2PbI_4$ (39 nm), due to the higher lattice rigidity of $(PEA)_2PbI_4$, which suppresses exciton-phonon coupling.[24] It has also been suggested that such an effect can further be enhanced by replacing MA with a larger FA as a cage cation in multilayered 2D perovskites.[21] Therefore, achieving a rigid, high symmetry and FA-based structure with minimum distortion at RT is expected to lead to significant improvement of exciton transport.

Here, we report for the first time, a distortion-free, FA-based Dion-Jacobson (FA DJ hereafter) 2D perovskites at RT using 3-(aminomethyl)piperidine (3AMP) as the A' site cation. The FA DJ 2D perovskite series (n=1 to 4) was synthesized using a novel approach of slow crystallization followed by hot extraction, which kinetically stabilizes the FA DJ 2D perovskite before it cools down to the undesired and thermodynamically stable, photo-inactive yellow or delta phase. Detailed crystallographic analysis shows that the DJ crystals exhibit the tetragonal P4/mmm symmetry throughout a large temperature range (0 to 150 ˚C). Structural characterizations reveal a Pb-I-Pb angle of 180˚ in FA DJ, inducing maximum overlapping between Pb s and I p orbitals, suggesting smaller exciton effective mass. Raman spectroscopy and solid-state magic-angle spinning nuclear magnetic resonance (MAS NMR) shows an extremely rigid octahedra network of FA DJ, reducing the exciton-phonon coupling, leading to faster exciton diffusion. Consequently, we measure an exciton diffusion length of 2.5 μm, and a diffusivity of 4.4 $cm^2s^{-1}$ of FA DJ are measured, both the largest among reported 2D perovskites, and also that of monolayer transition metal dichalcogenides, which exhibit exciton diffusion lengths of 1.5 μm.[25] To the best of our knowledge, this is the first demonstration of this maximum symmetry in multilayered 2D perovskite system at ambient conditions. From a technological perspective, This discovery marks a significant advancement in the development of next-generation optoelectronic devices, particularly excitonic systems such as nonlinear optical devices, quantum light sources, and photodetectors, as well as in the realization of technologically stable multijunction solar cells (e.g., Si/perovskite and perovskite–perovskite tandems).

## Result and discussion

**Hot-extraction synthesis and structural properties of FA DJ**

The FA-based 2D Dion–Jacobson (DJ) perovskite crystals were synthesized using a novel hot extraction method. As illustrated in Fig. 1a, formamidinium chloride (FACl) and lead oxide (PbO) powders (the rationale for using FACl instead of FAI is discussed in SI Section 1.1) were first mixed in hydroiodic acid (HI) solution and stirred at room temperature (25 °C). This rapidly produced a fresh yellow suspension, corresponding to the δ-phase of formamidinium lead iodide ($FAPbI_3$). As the hotplate temperature was increased to 230 °C under continuous stirring, the suspension gradually turned black, indicating the formation of α-phase $FAPbI_3$. To this hot α-$FAPbI_3$ suspension, a solution of 3-(aminomethyl)piperidine (3AMP), neutralized with hypophosphorous acid ($H_3PO_2$), was introduced. This led to the complete dissolution of black α-$FAPbI_3$, resulting in a clear yellow solution. The temperature was then gradually lowered to 120 °C, allowing for the slow crystallization of FA-based 2D perovskites. These DJ-type perovskites are characterized by an eclipsed conformation, with no offset between consecutive perovskite layers in the in-plane direction.

The synthesis of multilayered FA-based 2D perovskites (n > 2) has been particularly challenging, as it often relies on crystallization at RT, where the rapid nucleation of δ-$FAPbI_3$ occurs due to its low nucleation barrier at RT.[26,27] Consequently, the resulting crystals exhibit n = 1 or n = 2 phases but accompanied by δ-$FAPbI_3$ as a side product.[28–30] To address this issue, our hot extraction approach enables the isolation of FA-based DJ 2D crystals at elevated temperatures (>80 °C) over extended period, kinetically stabilizing them. This synthesis can further be understood by means of a schematic (non-quantitative) binary phase diagram, as shown in Fig. 1d and Fig. S1. The two sides of the phase diagram, (3AMP)PbI$_4$ and FAPbI$_3$,

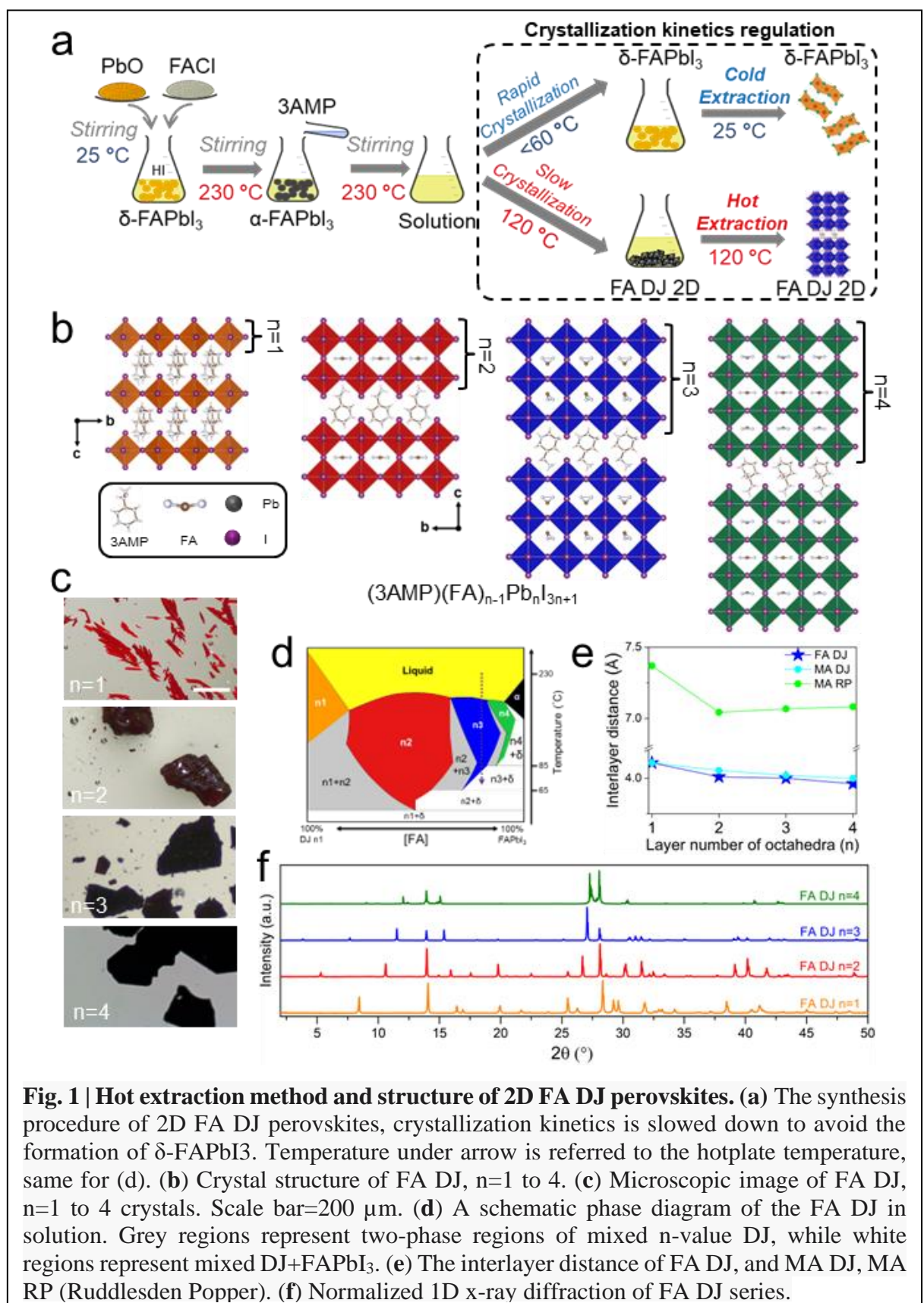

**Fig. 1 | Hot extraction method and structure of 2D FA DJ perovskites.** (**a**) The synthesis procedure of 2D FA DJ perovskites, crystallization kinetics is slowed down to avoid the formation of δ-FAPbI3. Temperature under arrow is referred to the hotplate temperature, same for (d). (**b**) Crystal structure of FA DJ, n=1 to 4. (**c**) Microscopic image of FA DJ, n=1 to 4 crystals. Scale bar=200 µm. (**d**) A schematic phase diagram of the FA DJ in solution. Grey regions represent two-phase regions of mixed n-value DJ, while white regions represent mixed DJ+FAPbI$_3$. (**e**) The interlayer distance of FA DJ, and MA DJ, MA RP (Ruddlesden Popper). (**f**) Normalized 1D x-ray diffraction of FA DJ series.

represent the two extreme stoichiometries of the FA DJ series when in an HI solution.

Intermediate stoichiometries can be written as $[(3AMP)PbI_4]_{1-x}[FAPbI_3]_x$ with $0<x<1$, or equivalently as $[(3AMP)PbI_4][FAPbI_3]_{n-1}$ where n is the number of octahedra layers. A schematic phase diagram can be generated by tracking which phases intersect with the convex hull of the Gibbs free energy landscape as the temperature rises and the free energy of the solution phase decreases. The method to construct this phase diagram is expanded upon in the SI. When using our FA DJ n=3 recipe, we move along the purple dashed line from the high-temperature liquid phase to the FA DJ n=3 phase. If the temperature is decreased further the solution moves into a two-phase region of both FA DJ n=3 and δ-$FAPbI_3$. This schematic phase diagram also rationalizes, why growing phase-pure FA DJ n=4 from solution is challenging, since the n=4 region is quite narrow and emerges at higher temperatures in comparison to n=3.

Fig. 1b shows the crystal structure of the $(3AMP)(FA)_{n-1}Pb_nI_{3n+1}$ series (FA DJ with n=1 to n=4), their detailed crystallographic data and structure refinement are listed in table S1. This homologous series is built from inserting 3AMP (3-(aminomethyl)piperidine) between layers of perovskites octahedra as the spacer, where the perovskite layer consists of formamidinium occupying the cage. All the analogs in this FA DJ series have a highly ordered perovskite layer, specifically n=2 to n=4 who take P4/mmm as the space group. This is the highest symmetry a 2D perovskite system can theoretically achieve.[31] The symmetry analysis was conducted by Quarti et al.[32] on a perfectly ordered 2D free-standing octahedra slab (considering only one horizontal plane and neglecting plane stacking along vertical direction) described by the P4/mmm layer group. The all-inorganic $Cs_2PbI_2Cl_2$, n=1 (Fig. S2) is so far the experimental 2D lead-halide perovskite structure exhibiting the highest symmetry.[33] Its I4/mmm tetragonal centered space group, corresponds to a Ruddlesden-Popper (RP) ordering, with a (1/2,1/2) in-plane shift between successive layers along the stacking axis.[33,34] Our FA DJ series is the first multiple-layer 2D perovskite series that takes the highest P4/mmm symmetry. For n=1 crystals, two different polymorphs were found, with a monoclinic one previously reported[13] exhibiting well defined motifs for the organic cations in the interlayer and another tetragonal, which shows similar space group symmetry and cation disorder as the n=2 to n=4 structures (Fig. S3). When a crystal initially in the monoclinic n=1 phase was heated up to 190 °C, it transforms into the tetragonal phase through a first order phase transition and stabilized even cooled down later (Fig. S3d). In this paper we mainly discuss the tetragonal n=1 phase unless specified otherwise. Fig. 1c shows the microscopic image of the FA DJ 2D crystals, from n=1 to n=4 the color changes from fresh-red to dark-red and finally black.

The in-plane lattice parameter of FA DJ series and the one of α-FAPbI$_3$ is plotted in Fig. S4. After the introduction of FA into the structure, a dramatic increase of the in-plane lattice parameter from n=1 to n=2 is observed, and then progressively approaching the parameter of 3D α-FAPbI$_3$ when n further increases. Here to compare the lattice parameters between structures exhibiting different symmetries, an "effective in-plane lattice parameter" is considered (Table S2), which is essentially the width of a PbI$_6$ octahedron. On the other hand, the in-plane lattice parameter of the monoclinic n=1 structure is larger than for the tetragonal n=1 one because the conformation of the interlayer cation breaks the local symmetry. The 1D X-ray diffraction patterns on crystals of the FA DJ series are shown in Fig. 1f and Fig. S5, where the low angle peaks are apparent and well-defined for n=1 to n=4. For n=4 the intensity of low angle peaks is weak and could be better observed with log-scaled intensity (Fig. S6). These characteristics match well with the previously reported 2D perovskite structures.[12,35,36]

The interlayer distance of FA DJ 2D perovskites is the shortest compared to reported 2D perovskites (Fig. 1e).[13,37] For MA-based 2D perovskites, the DJ type (e.g. MA 3AMP 2D)[13] is generally much shorter than Ruddlesden Popper (RP) type (e.g. MA butylammonium (BA) 2D)[12] because there is one layer of divalent spacer cation in DJ versus two layers of monovalent spacer in RP. For RP 2D perovskites, the interlayer distance is the distance between two (010) planes across apical iodides (Fig. S7). For DJ 2D perovskites, the interlayer distance is essentially the distance between two apical iodides since the perovskite layers stack exactly on top of each other. While the interlayer distance of MA DJ is already one of the shortest (~4.0 Å for n=4), for FA DJ it is even shorter (3.96 Å for n=4).

**Perovskite layers with no octahedral distortions and optical properties**

Next, we investigated the correlations between the maximal symmetry observed for the new 2D tetragonal FA DJ structures and their optical properties. As shown in fig. 2a and 2b, almost all the reported 2D perovskite structures exhibit either in-plane distortion or out-of-plane distortion, or both at ambient conditions. However, FA DJ series shows a linear structure with no distortion along both in-plane and out-of-plane directions, which has never been reported in any hybrid multilayered (n>1) 2D perovskites.

Fig. 2c shows the average equatorial Pb-I-Pb angle of 3 types of 2D perovskites in fig. 2a and 2b and α-phase FAPbI$_3$. The average equatorial Pb-I-Pb angles of both MA BA and MA DJ series are smaller than the average Pb-I-Pb angle in 3D FAPbI$_3$ (180°). After an initial steep variation from n=1 to n=2, they slowly change as a function of n-value toward values observed for the 3D reference MAPbI$_3$, as the effect of organic cation on the inorganic slabs progressively weakens.[12,13] However, the FA DJ series shows for all the compounds with n>1 (n=2,3,4), equatorial Pb-I-Pb angles of 180° equivalent to the average Pb-I-Pb angle in the α-phase FAPbI$_3$. These observations in crystal structure are supported by the analysis of $^{207}$Pb and $^1$H solid-state NMR spectra (ssNMR, Fig. 2d-e). The identical $^{207}$Pb shifts at 1560 ppm observed for the FA DJ series (n=2, 3) and 3D FAPbI$_3$ correspond to equatorial Pb-I-Pb angles of 180°. In contrast, the n=1 DJ phase exhibits a shift in the $^{207}$Pb chemical shift to 1230 ppm, indicative of tilted Pb-I-Pb angles (166°). Analysis of $^1$H NMR chemical shifts further supports the role of non-covalent interactions at the spacer/perovskite interface in stabilizing the perovskite structure. This is evident from the hydrogen bonding-dependent chemical shifts of NH, NH$_2$, and NH$_3$ moieties in AMP and FA cations, which are identical in n=2, 3 and 3D FAPbI$_3$, but differ in n=1. Resonances at 7.4 ppm and 8.1 ppm arise from the identical local chemical environments of CH and NH$_2$ sites in FA cations in n= 2, 3 FA DJ phases and α-FAPbI$_3$. The broad distribution of $^1$H peaks in the 1–5 ppm range can be attributed to the -NCH$_2$-, -CH$_2$-, and CH moieties of cyclic piperidinium, while the -NH$_2^+$ and NH sites produce peaks at 7.1 and 7.4 ppm (black dots), respectively. These peaks differ between n=1 and n=2, 3 FA DJ phases, with $\Delta\delta(NH_2^+)= 0.2$ ppm (vertical dashed line). The distinct chemical shifts indicate variations in non-covalent interactions at the spacer/perovskite interface, highlighting their crucial role in maintaining structural integrity and a high degree of local order within the inorganic slabs. Together, these results indicate that the Pb-I-Pb bond is fully stretched thanks to the introduction of FA into the cage, and the stress induced by the 3AMP organic spacer cation is negligible by comparison to the negative pressure on the perovskite backbone imposed by FA.

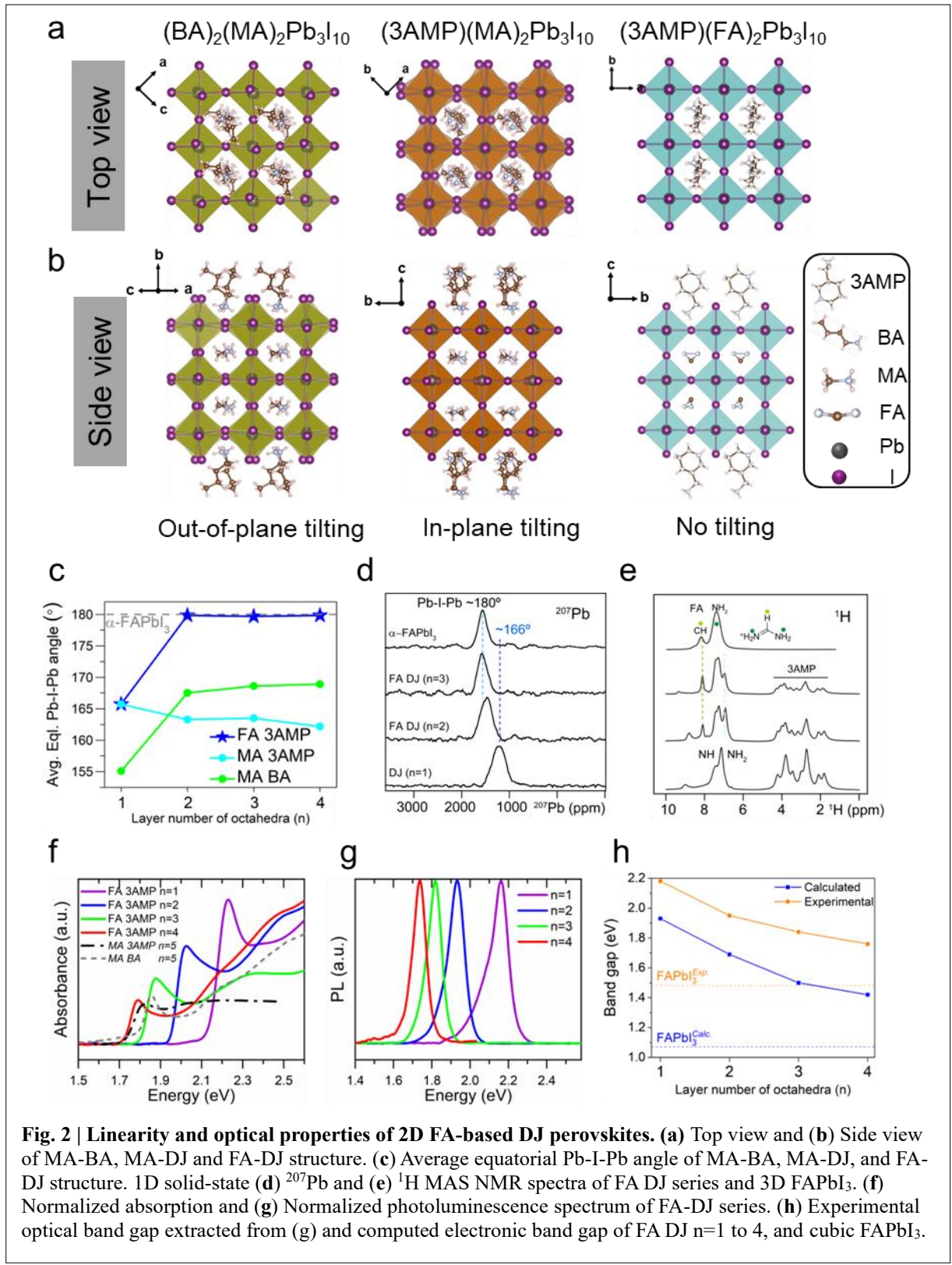

**Fig. 2 | Linearity and optical properties of 2D FA-based DJ perovskites. (a)** Top view and **(b)** Side view of MA-BA, MA-DJ and FA-DJ structure. **(c)** Average equatorial Pb-I-Pb angle of MA-BA, MA-DJ, and FA-DJ structure. 1D solid-state **(d)** $^{207}$Pb and **(e)** $^1$H MAS NMR spectra of FA DJ series and 3D FAPbI$_3$. **(f)** Normalized absorption and **(g)** Normalized photoluminescence spectrum of FA-DJ series. **(h)** Experimental optical band gap extracted from (g) and computed electronic band gap of FA DJ n=1 to 4, and cubic FAPbI$_3$.

The high structural ordering of the perovskite backbone on the average plays a crucial role in determining the electronic properties. It is known for layered perovskites that, as Pb−I−Pb bond angles increased to 180°, overlapping between Pb s and I p orbitals increases, pushing up the valence band maximum and finally reducing the electronic bandgap.[13,38–40] When compared to MA RP and MA DJ (Fig. 2f), the FA DJ series shows indeed a systematic narrower band gap for a given n value, bringing the exciton peak of FA DJ n=4 close to the exciton of n=5 of the MA analogs. From the steady-state photoluminescence (PL, Fig. 2g) maxima, we extracted the optical band gaps of the FA DJ compounds, as shown in Fig. 2h. Notably, the optical band gap of 1.76 eV for n = 4 is the smallest reported so far among all known pure 2D perovskites. Additionally, our calculated electronic band gaps for the FA DJ series, based on first-principles calculations, clearly confirm the effect of quantum confinement in this family of layered materials. We employ the DSH hybrid functional (see methods) that is known to give excellent agreement with experimental band-gaps of halide perovskites.[41] Our calculations nicely capture the changes of both the bandgap and the dielectric constant when going from n=1 to n=4 (Table

S4). As shown in Fig. 2h, the calculations underestimate the band gaps, partly due to not including self-energy corrections related to the anharmonic lattice dynamics[42,43]. We estimate the order of magnitude of the effect on FAPbI$_3$ by calculating a band gap increase of 240 meV. In supplementary Fig. S8 we scissor shift all bandgaps by this value.

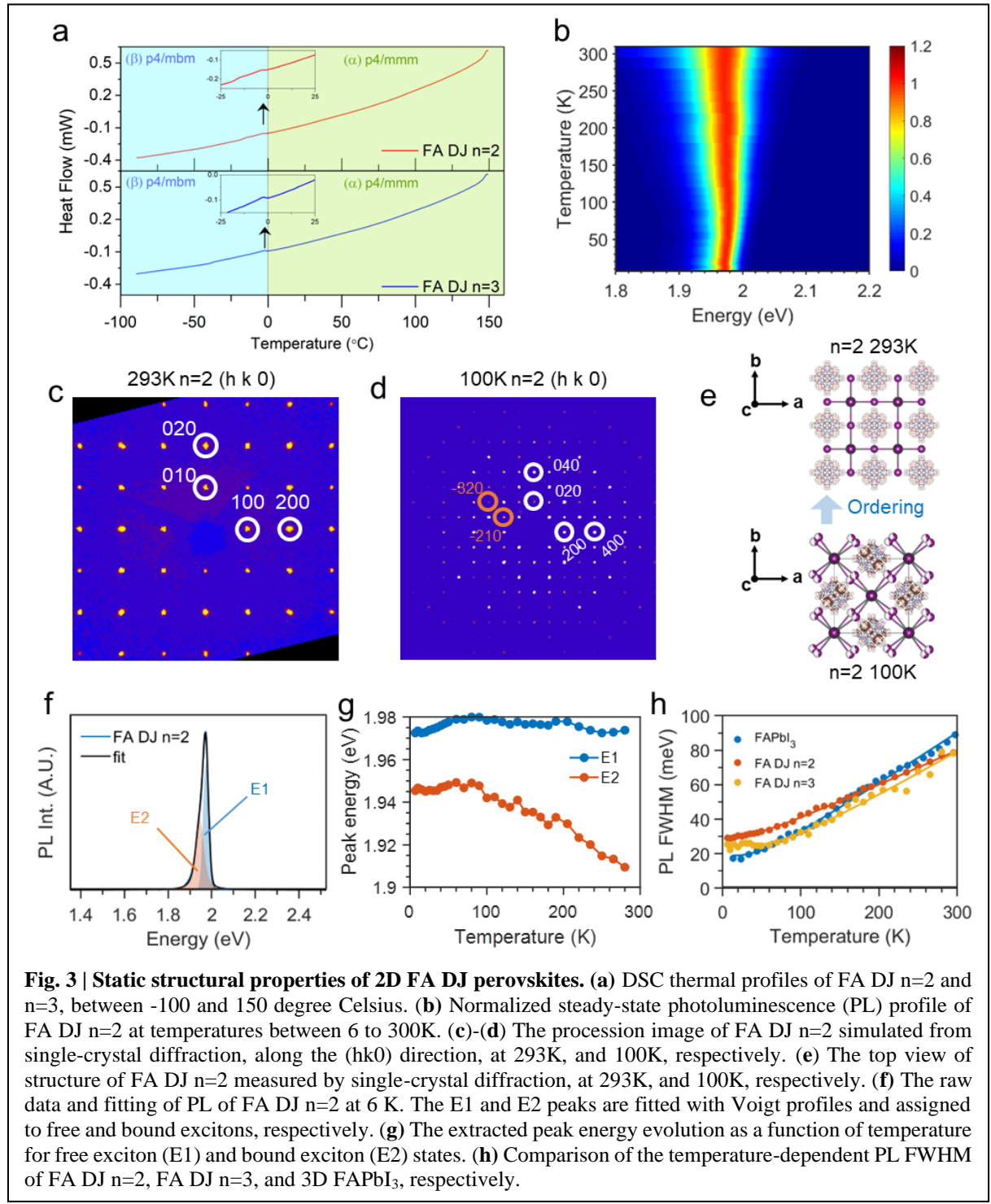

**Fig. 3 | Static structural properties of 2D FA DJ perovskites. (a)** DSC thermal profiles of FA DJ n=2 and n=3, between -100 and 150 degree Celsius. **(b)** Normalized steady-state photoluminescence (PL) profile of FA DJ n=2 at temperatures between 6 to 300K. **(c)**-**(d)** The procession image of FA DJ n=2 simulated from single-crystal diffraction, along the (hk0) direction, at 293K, and 100K, respectively. **(e)** The top view of structure of FA DJ n=2 measured by single-crystal diffraction, at 293K, and 100K, respectively. **(f)** The raw data and fitting of PL of FA DJ n=2 at 6 K. The E1 and E2 peaks are fitted with Voigt profiles and assigned to free and bound excitons, respectively. **(g)** The extracted peak energy evolution as a function of temperature for free exciton (E1) and bound exciton (E2) states. **(h)** Comparison of the temperature-dependent PL FWHM of FA DJ n=2, FA DJ n=3, and 3D FAPbI$_3$, respectively.

**Static structural properties of 2D FA DJ perovskites**

Antiferrodistorsive phase transitions from the high temperature maximally symmetrized cubic phase are commonly observed in 3D halide perovskites. Similar phase transitions for 2D FA DJ perovskites were first evidenced by using Differential Scanning Calorimetry (DSC) (Fig. 3a). Both n=2 and n=3 exhibit a symmetry reduction from α-tetragonal P4/mmm to β-tetragonal P4/mbm at 0 °C (273 K), corresponding to a second order phase transition. Similar phase transitions at low temperature was also observed in 3D $FAPbI_3$ (and $CsPbI_3$), which exhibited a second order, cubic-to-tetragonal (α-to-β) phase transition at 280 K.[44] It should be noted indeed that a classical symmetry analysis combined with Landau theory of phase transition predicted a continuous transition (second-order or weakly first order phase transition) for both Pm-3m to P4/mbm ($FAPbI_3$) and P4/mmm to P4/mbm (FA DJ series).[45]

From XRD procession images (Fig. 3c and 3d), systematic superlattice reflections are observed at 100K along the (h k 0) plane (Fig. 3d, highlighted in orange) as compared to the room temperature P4/mmm structure (Fig. 3c), which necessitates an expansion of the unit cell by a factor of √2 in both the a and b directions. This new unit cell meets all the systematic absence conditions for the P4/mbm space group. As shown in Fig. 3e, upon cooling down to 100K, the Pb-I-Pb bond angle is bent from 180° to 166º in the P4/mbm structure, with all the distortion occurring in the plane of the $PbI_4$ layers. Analogous transitions were observed for n = 3 and 4 compounds (Fig. S9 to S11) where the equatorial Pb-I-Pb angles transition from 180º to 164.21(5)º and 162.58(7)º, respectively (Crystallographic Information File (Cif) of n=2,3,4 at 100K can be found in SI). Phase transitions where the angles straighten to 180º at higher temperatures are well reported and documented for halide perovskites such as $CsPbI_3$,[4,46,47] and in 2D perovskites the angle straightening as a function of temperature has been observed in prototypical examples such as the $(C_nH_{2n+4}N)_2PbI_4$ (n = 4-6) series.[48]

As shown in Fig. 3b (n=2) and Fig. S12 (n=3), the exciton PL in both cases exhibit a gradual energy evolution from 6K to 300K, consistent with a continuous phase transition from a low-temperature β-tetragonal to high temperature α-tetragonal phase in FA DJ series. At very low temperatures (6 - 100K), the PL demonstrates a further subtle blue shift in both n=2 and n=3 perovskites, which also agrees with experimental results for the $FAPbI_3$ films in the same temperature range where the orthorhombic g phase is expected to be stable.[49] No sub-bandgap spectroscopic signatures are detected in the PL or reflectance spectra (Fig. S12 and S13), indicating ultralow densities of deep traps and excitonic transitions in FA DJ perovskites. As shown in Fig. 3f, the exciton PL at 6K demonstrates an asymmetric line shape contributed by two emission peaks, which are assigned to free exciton (blue, E1) and bound exciton states that shall be associated with shallow defects (Orange, E2) respectively. As displayed in Fig. 3g, the temperature-dependent energy splitting between the two peaks further confirms the assignment of bound exciton states instead of phonon replica.[50] Fig. 3h compares the extracted PL linewidths of n=2, n=3, and 3D $FAPbI_3$ as a function of temperature. Both FA DJs show ~3 times increase of FWHM from 6K to 300K (30meV to ~80 meV), with similar trends of thermal broadening as 3D $FAPbI_3$ thin films. Quantitative analysis of temperature-dependent linewidth was performed assuming an effective electron-phonon coupling mechanism for optical phonons. (Detailed methods and fitting parameters are listed in SI discussion 1.4 and Table S3) We extracted the effective longitudinal optical (LO) phonon energy to be ~10meV for n=2 and 20meV for n=3, showing similar reported values in both 3D $FAPbI_3$ thin films and nanocrystals,[49,51] consistent with electron-phonon coupling mechanisms in the multi-layered 2D DJ perovskites similar to the ones in 3D $FAPbI_3$.

**Lattice dynamics of 2D FA DJ perovskites**

Raman scattering provides a first indication about the nature of lattice dynamics of the new FA DJ series (Fig. 4a). At room temperature, the lattice dynamics of the FA DJ compounds for n>1 exhibit mainly relaxation-like spectral continuum in their quasi-elastic scattering signatures in the 0-100 cm$^{-1}$ range, which was commonly observed in 3D perovskite materials[52], including 3D α-$FAPbI_3$[53]. In contrast, the Raman scattering signature of the n=1 DJ compound, which does not contain any FA cations, exhibits well-defined low-frequency peaks in the 0-60 cm-1 range. For the n>1 compounds, the only clear Raman scattering resonance still observed is located at about 120cm$^{-1}$ (15 meV) corresponding to the group of highest energy longitudinal optical (LO) phonons, which is consistent with the effective optical phonon frequency determined from the PL broadening variation as a function of temperature (Fig. 3h). Interestingly, persistence of broadened high-frequency LO phonon modes, accompanying an overdamped low-frequency spectral continuum, have been similarly observed in Raman scattering of 3D bromide and iodide perovskites.[50,51]

To get a better understanding about the microscopic origin of the dynamical disorder, we first analyzed the anisotropic thermal displacement parameters extracted from the X-ray diffraction analysis of the FA DJ structures (Fig. 4b and Fig. 4c). X-ray diffraction is very sensitive to contributions from the Pb and I atoms, and here we focused on I atoms, which are the most affected by octahedra rotations. As shown in Fig. 4c, the apical iodine in the n=1 compound has an in-plane displacement parameter different from the ones in other FA DJ compounds and α-FAPbI$_3$. Internal iodides behave similarly in n>1 FA DJ compounds and α-FAPbI$_3$. The main difference comes from apical iodide atoms, which exhibit a very large thermal disorder. These atoms are indeed in direct contact with 3AMP interlayer cations, which are disordered in these structures.

From the XRD displacement parameters (Fig. 4c), we know that the apical iodides located at the edges in n>1 FA DJ compounds and in direct contact with FA and 3AMP cations exhibit a higher degree of disorder than the iodides located in the inner layers of the perovskite backbone. ssNMR analysis was conducted to gain insight into the local structures and dynamics. The nuclear spin-lattice relaxation ($T_1$) times (Fig. 4d-e) of FA$^+$ and 3AMP spacer cations are sensitive to the local fluctuations in the inner and apical iodine sites of the PbI$_6$ octahedra in 2D FA DJ phases. Although the resonances of FA cations are identical in the FA DJ and α-FAPbI$_3$ phases as previously shown in Fig. 2e, their site-specific $T_1$ values increase as n increases: 36.8 s for α–FAPbI$_3$, 2.4 s in FA DJ (n=3) and 1.6 s in FA DJ (n=2). The observations are corroborated by the reduced $T_1$ values of NH sites cations (3AMP), which are 1.9 s in FA DJ (n=3) and 1.5 s in FA DJ (n=2). The disorder of apical iodides at the edges thus correlates with the presence of 3AMP cations in the interlayer. Using n>1 FA-based 2D DJ compounds almost lattice matched with α-FAPbI$_3$, has a beneficial effect on the diffusion length of excitons by comparison to known 2D perovskites. This is attributed to the presence of FA as a cage cation (vide infra). But the presence of larger cations in the interlayer with different properties than FA and correlated with a larger disorder of the apical iodide atoms, appears to set an upper limit to the propagation of free excitons. This interface effect is mitigated by enlarging the inner layers.

In Fig. 4f and 4g, the calculated phonon spectral functions in the frequency range 0 – 10 meV of cubic FAPbI$_3$ and FA DJ n=2 are compared. In this range, the spectrum is dominated by vibrations of the inorganic network. Polymorphism in both cases induce coupled optical vibrations, reducing the phonon correlation length and lifetimes. The enhanced broadening observed for FA DJ n=2 was related to quantum confinement and the disconnection of the perovskite backbone along the stacking axis, leading essentially to weaker elastic constants and surface-like optical modes. Nonetheless, the phonon spectral function calculated for FA DJ

n=2 exhibits similar phonon frequencies and quasiparticle peak positions to those calculated for $FAPbI_3$. It is expected that these features will converge in the bulk limit.

In Fig. S21, we show the valence band maximum (VBM) and conduction band minimum (CBM) density of states (DOS) of polymorphous FA DJ n=2 and $FAPbI_3$ calculated for 300 K. Our results show that both the VBM and CBM of FA DJ n=2 exhibit a similar band broadening than the corresponding single electronic states of $FAPbI_3$. Notice that the PL linewidth (Fig. 3h) is not directly related to the smearing of the electronic density of states but to the effect of the imaginary part of the electron-phonon self-energy on resonant excitonic states.[54] From the real part of the electron-phonon self-energy, our calculations for the phonon-induced band gap renormalization $[\Delta\varepsilon_g(T)]$ demonstrate a stronger electron-phonon coupling effect in $FAPbI_3$. In particular, our calculations yield a band gap opening of $\Delta\varepsilon_g(300\ K) = 120$ meV for $FAPbI_3$, while $\Delta\varepsilon_g(300\ K) = 33$ meV for FA DJ n=2, i.e. 4 times smaller.

Finally, we performed real-space PL measurements to directly correlate the degrees of octahedra distortion with the in-plane exciton diffusion lengths of FA DJ n=3 (Fig. 4h, Fig. S22). The real-space landscape shows a PL gradient where the maximum intensity is formed at the excitation position and spreads radially with decreasing intensity (Fig. 4h). Radiative recombination from free carriers was spectrally filtered, thus the PL intensity maps show the in-plane transport of excitons near the band edge (1.82 eV) of FA DJ n=3. The radial symmetry of the diffusion indicates surface homogeneity and unconfined exciton migration within the plane. We also do not observe significant line shape changes for power dependent profiles of the n=3 exciton diffusion (Fig S20a). To quantify the diffusion length of the FA DJ perovskites, the PL intensity was radially averaged about the excitation center, normalized, then bound by a fitting function for several powers (Fig. S23). This fit is given by the asymptotic function, $e^{-r/L_D}/\sqrt{r/L_D}$, where $r$ is the displacement from the center of excitation and the diffusion length, $L_D$, is the fit parameter (SI section 1.6).[55,56] The calculated lower and upper limits for the n=3 diffusion lengths were 1.8 µm and 2.5 µm respectively, longest so far compared to previous reports (Fig.4i).[24,57] We observe fluence independence of the exciton diffusion for powers ≤1483 µW. At higher laser powers (≥ 2500 µW) exciton-exciton annihilation occurs near $r = 0$ and artificially increases the exciton diffusion.[58] These higher powers were omitted due to the unreliability of their diffusion lengths. FA DJ n=2 demonstrated smaller diffusion lengths of 0.83 – 1.5 µm (Fig S20b), an expected behavior due to n-dependence typically observed in 2D perovskites.[24] Sizeable exciton diffusion lengths are necessary in order to ensure a proper exciton dissociation at the perovskite layer edges in solar cells containing thick and vertically oriented 2D layers.[59,60]

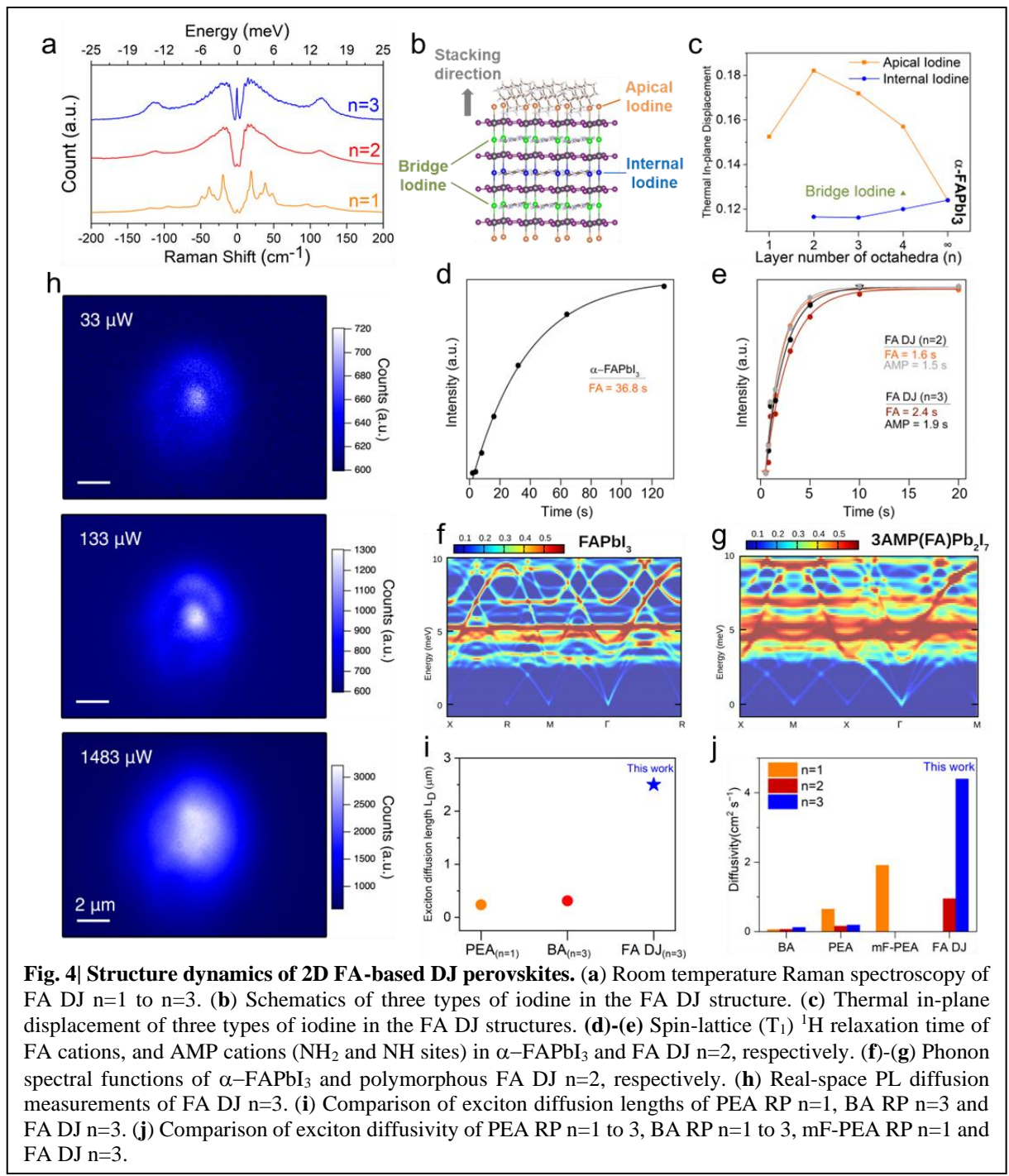

**Fig. 4| Structure dynamics of 2D FA-based DJ perovskites.** (**a**) Room temperature Raman spectroscopy of FA DJ n=1 to n=3. (**b**) Schematics of three types of iodine in the FA DJ structure. (**c**) Thermal in-plane displacement of three types of iodine in the FA DJ structures. (**d**)-(**e**) Spin-lattice ($T_1$) $^1$H relaxation time of FA cations, and AMP cations ($NH_2$ and NH sites) in α−$FAPbI_3$ and FA DJ n=2, respectively. (**f**)-(**g**) Phonon spectral functions of α−$FAPbI_3$ and polymorphous FA DJ n=2, respectively. (**h**) Real-space PL diffusion measurements of FA DJ n=3. (**i**) Comparison of exciton diffusion lengths of PEA RP n=1, BA RP n=3 and FA DJ n=3. (**j**) Comparison of exciton diffusivity of PEA RP n=1 to 3, BA RP n=1 to 3, mF-PEA RP n=1 and FA DJ n=3.

In order to perform a physical comparison with previous studies on 2D multilayered RP perovskites, with PEA and BA cations in the interlayer (($BA)_2(MA)_{n-1}Pb_nI_{3n+1}$ and ($PEA)_2(MA)_{n-1}Pb_nI_{3n+1}$, with n=1-3 (Fig. 4j),[21,61] we estimated the exciton diffusivity for FA DJ n=2 and n=3 (SI section 1.6, Fig. S24). The estimated diffusivity for DJ FA n=3 ($D_X \sim 4.4 cm^2 s^{-1}$) is the largest obtained so far for 2D perovskites[62], sizeably larger than the previous record value reported for (mF-PEA)$_2$PbI$_4$ ($D_X \sim 1.9 cm^2 s^{-1}$)[61]. A reliable theoretical framework and microscopic interpretation of the intrinsic exciton diffusivity in 2D multilayered perovskites does not exist yet. It has thus been proposed to correlate the observed exciton diffusivity to various experimental results, related to structural distortions including in-plane and out of plane perovskite octahedra tilt angles, exciton–phonon coupling strength estimated from temperature dependent PL broadening and average atomic displacements.[21,61] Small values for the last two parameters were interpreted as signatures of a reduced exciton-phonon coupling strength and a large lattice stiffness due to the presence of a rigid cation. This was the main explanation given for the sizeably larger exciton diffusivities in PEA RP n=1 2D perovskites by comparison to (BA)$_2$PbI$_4$. Following this proposition, large values of the PL broadening and atomic displacement parameters in the present work may indicate that the exciton coupling strengths are much larger for FA DJ perovskites ($\Gamma_{LO}$ =50meV for FA DJ n=3 in the present work, compared to 14.96 meV for PEA RP n=3 and 22.25 meV for BA RP n=3). However, as shown by our simulations, polymorphism is important in FA DJ perovskites. This effect is known to lead to a smearing of both vibrational and electronic dispersion.[42] Therefore, PL broadening and atomic displacement parameters are not expected to be solely related to the exciton coupling strength but are additionally increased by polymorphism.

These studies directly show that impact of replacing MA cations by FA cations in enhancing exciton diffusivities, which is attributed to the increased lattice stiffness of the structures containing FA.[21] Next, it is also necessary to consider the effect of the increase of quantum well thickness (n). It is reported in previous studies that BA-based 2D perovskites undergo a clear increase of exciton diffusivities from n=1 to n=3, which correlates to the reduction of the octahedra tilt angles, when the n value approaches 3D perovskites. By comparison, the PEA RP perovskites exhibit the opposite trend, with larger octahedra tilt angles for the n=3 compound than for the BA-based one. This second aspect is also important for the n=3 FA DJ compound reported in the present work, which exhibits both a large exciton diffusivity and vanishing octahedra tilt angles.

It can be inferred from this study that the beneficial effect of replacing MA cations with FA cations in enhancing exciton diffusivities is attributed to the increased lattice stiffness of the structures containing FA.[21] Next, it is also necessary to consider the effect of the increase of quantum well thickness (n). It has been reported in previous studies that BA-based 2D perovskites exhibit a significant increase in exciton diffusivity from n=1 to n=3, which correlates with the reduction of octahedra tilt angles as the n value approaches that of 3D perovskites. By comparison, the PEA RP perovskites exhibit the opposite trend, with larger octahedra tilt angles for the n=3 compound than for the BA-based one. This second aspect is also important for the n=3 FA DJ compound reported in the present work, which exhibits both a large exciton diffusivity and vanishing octahedra tilt angles. In summary, our work paves the path for the design of soft semiconductors with high symmetry through the use of appropriate cations and directly demonstrates its impact on physical properties such as exciton diffusion lengths, a property, which is critical for the design of optoelectronic and photonic devices of the future.

**Data availability:** Crystallographic data for the structures reported in this article have been deposited at the Cambridge Crystallographic Data Centre, under deposition numbers CCDC 2433609-2433615. All the data supporting the findings of this study are available in the Article and Supplementary Information. Any additional data is available from the corresponding authors upon reasonable request. Source data are provided with this paper.

**Acknowledgments:** J.H. acknowledges the financial support from the China Scholarships Council (No. 202107990007). M.Z. acknowledges funding from the European Union's Horizon 2020 research and innovation program under the Marie Skłodowska-Curie Grant Agreement No. 899546. G.V. acknowledge funding from the Agence Nationale pour la Recherche through the CPJ program and the SURFIN project (ANR-23-CE09-0001). M.Y.S. was supported by the donors of ACS Petroleum Research Fund under Grant No. 65743-ND6. C.W. and G.N.M.R thank EU H2020 (No. 795091) and INFRANALYTICS FR-2054 CNRS. J.E. acknowledges financial support from the Institut Universitaire de France. S.J.H. acknowledges support from NSF Grant No. HRD-2112550 (Phase II CREST Center IDEALS). S.S and B.Z acknowledge support from NSF/EPSCoR RII Track-1: EQUATE under Award No. OIA-2044049. Y.G. acknowledges support from NSF under Award No. DMR-2339721.

**Author contributions:** A.D.M. conceived the idea. J.H. developed the synthetic method, conducted the experiments. J.F. performed the single crystal X-Ray diffraction and refinement of the crystal structure under the guidance of M. G. K. J.H. performed 1D-XRD measurements with the help of I.M. H.Z. and J.H. performed optical characterizations. S.J.H. performed real-space diffusion measurements, corresponding diffusion length calculations and analysis under the supervision of M.Y.S.. C.W. and G.N.M.R performed ssNMR measurements and analysis. S.S and B.Z performed low frequency Raman spectroscopy measurements under the guidance of Y.G. M.Z. and G.V. performed the calculations and analyzed the computational results. J.H., J.F. H.Z. performed data analysis with guidance from C.K.,

M.K., J.E., and A. D. M.. J.H. wrote the manuscript with input from everyone. All authors read the manuscript and agree to its contents, and all data are reported in the main text and supplemental materials.

**Ethics declarations**

**Competing interests:** Rice University has filed a patent for a method of fabricating the FA DJ 2D perovskite. The other authors declare no competing interests.

Supplementary Information for

# Two-dimensional perovskites with maximum symmetry enable exciton diffusion length exceeding 2 micrometers

Jin Hou[1, #], Jared Fletcher[2, #], Siedah J. Hall[3,4], Hao Zhang[5,6], Marios Zacharias[7], George Volonakis[8], Claire Welton[9], Faiz Mandani[5], Isaac Metcalf[1], Shuo Sun[10], Bo Zhang[10], Yinsheng Guo[10], G. N. Manjunatha Reddy[9], Claudine Katan[8], Jacky Even[7], Matthew Y. Sfeir[3,4], Mercouri G. Kanatzidis[2*] and Aditya D. Mohite[1,5*]

[1]**Department of Materials Science and NanoEngineering, Rice University, Houston, Texas 77005, USA.**
[2]**Department of Chemistry and Department of Materials Science and Engineering, Northwestern University, Evanston, Illinois 60208, USA.**
[3]**Photonics Initiative, Advanced Science Research Center, City University of New York, New York, 10031, USA**
[4]**Department of Physics, The Graduate Center, City University of New York, New York, 10016, USA**
[5]**Department of Chemical and Biomolecular Engineering, Rice University, Houston, Texas 77005, USA.**
[6]**Applied Physics Graduate Program, Smalley-Curl Institute, Rice University, Houston, TX, 77005, USA.**
[7]**Univ Rennes, INSA Rennes, CNRS, Institut FOTON - UMR 6082, 35708 Rennes, France.**
[8]**Univ Rennes, ENSCR, INSA Rennes, CNRS, ISCR (Institut des Sciences Chimiques de Rennes) - UMR 6226, F-35000 Rennes, France.**
[9]**University of Lille, CNRS, Centrale Lille Institute, Univ. Artois, UMR 8181−UCCS− Unité de Catalyse et Chimie du Solide, F-59000, Lille, France.**
[10]**Department of Chemistry, University of Nebraska-Lincoln, Lincoln, NE 68588**

#Contributed equally

*Correspondence: m-kanatzidis@northwestern.edu, adm4@rice.edu

**Table of Contents**

# 1. Supplementary Text

## 1.1 Discussion of the Formamidinium (FA) source in the precursors.

Formamidinium Iodide (FAI) could also be used but doesn't work as well as FACl as it sometimes leads to extra formation of δ-phase $FAPbI_3$. Because there is a large excess amount of halide from hydrohalic acid in the solution, the content of halogen in the final perovskites product is only determined by the halide in the hydrohalic acid (e.g. HI acid solution will lead to iodine based perovskites, HBr acid solution will lead to bromine based perovskites, and HI/HBr mixed solution will lead to I/Br perovskite[1]). Here, as long as HI is used as the solvent, it will be the exclusive halide source, and the choice of halogen in the FA source (either FAI, FACl, FABr) or metal source (either $PbI_2$, $PbCl_2$, $PbBr_2$ or PbO which will become $PbX_2$, X as a halogen after dissolving) won't influence the final composition.[2]

## 1.2 Discussion of crystallization kinetics, temperature, and yield of the FA Dion-Jacobson (FA DJ) synthesis.

The crystallization temperature can be further lowered to 100 ˚C (referring to the temperature of hotplate, throughout this section, unless specified otherwise) to increase the yield, but also with more risk of having more $FAPbI_3$ δ-phase in the product as impurity. If the temperature is lower below 100 ˚C, for example room temperature (25 ˚C), the yellow $FAPbI_3$ δ-phase will dominate the crystallization. We think kinetically FA Dion-Jacobson (FA DJ) 2D is much slower in terms of crystallization compared to yellow $FAPbI_3$ δ-phase, which is why high temperatures, longer time strategy is utilized here to have phase-pure $(3AMP)(FA)_{n-1}Pb_nI_{3n+1}$ (FA DJ 2D hereafter, 3AMP=3-(aminomethyl)piperidine). However, as we must crystallize at higher temperatures, we sacrifice the yield since at higher temperatures the solubility of FA DJ 2D in this aqueous solution is higher, leaving a fair amount of precursors still dissolving in the solution. This is also why we use such a large scale of synthesis for these materials.

**1.3 Discussion of synthesis of FA DJ powders.**

The synthesis of n=1 to n=3 powders are highly reproducible, but we do notice that, when changing between precursor chemicals (for example, from different batches, or from different vendors), or between seasons (there is large variation of humidity in Houston, where synthesis has been performed), occasionally the recipe needs to be slightly adjusted to get phase-pure crystals. The adjustment is simply changing the FACl to 3-(aminomethyl)piperidine (3AMP) (cage cation to spacer cation) ratio. An elevated ratio will lead to higher n and lowering the ratio will lead to lower n. For example, if a n=2 and n=3 mixture is obtained when targeting n=3, slightly adding more FACl and less 3AMP will lead to pure n=3. Based on our experience, this rule is generally applicable to all the 2D halide perovskite synthesis and very useful practically.

For the synthesis of the powder form n=4, this rule holds true; however, there are thermodynamic factors that should be considered. Specifically, when keep rising the ratio of FACl to 3AMP, the formation of n=4 comes at the expense of side product, namely, yellow $FAPbI_3$ δ-phase, and the amount of $FAPbI_3$ δ-phase increases as we further raise the FACl: 3AMP ratio. We think in terms of thermodynamics, the enthalpy of formation[3] (an indicator of whether a 2D perovskite is favorable or not) of $FAPbI_3$ δ-phase will be between FA DJ n=3 and n=4. Therefore, n=4 will be less favorable compared to $FAPbI_3$ δ-phase. However, we could make small but pure n=4 single crystal using our previously reported KCSC method[4], as it enabled us to grow a unfavorable phase over long time via transformation from lower n to higher n.

**1.4 Discussion of the temperature-dependent linewidth and exciton-phonon couplings in 2D FA DJ perovskites.**

The temperature-dependent PL of FA DJ n=2 and n=3 samples are analyzed by fitting the PL spectra shape with Voigt profiles and extracting their FWHMs as a function of temperature Γ(T). The extracted linewidths are plotted in Fig. 3e. Based on the gradient of the PL linewidths near T = 0K, we estimate that the acoustic phonon contribution to the exciton linewidth broadening is minimal ($\gamma_{ac} <$ 70 μeV/K for n=2, and $\gamma_{ac} <$ 10 μeV/K for n=3, extracted by taking the gradient $d\Gamma/dT$ at T = 0K).[5] Therefore, we neglect the exciton-acoustic phonon coupling term (behaves linearly with temperature), and fit the temperature-dependent PL broadening by only considering the inhomogeneous broadening and exciton-longitudinal optical (LO) phonon coupling, which is given by:

$$\Gamma(T) = \Gamma_0 + \Gamma_{LO} \frac{1}{e^{E_{LO}/k_B T} - 1}$$

Here, $\Gamma_0$ represent the zero-temperature linewidth originating from exciton inhomogeneous broadening, $\Gamma_{LO}$ represents the strength of exciton-LO phonon coupling, and $E_{LO}$ is the corresponding LO phonon energy. The fitted parameters are shown in Table S2. The extracted LO phonon energies are for ~8meV n=2 and ~18meV for n=3. It suggests that the strength of the electron-phonon coupling in FA DJ n=2 and n=3 is not significantly different than the 3D FAPbI$_3$ phase.

## 1.5 Discussion of the Solid-state Magic-Angle Spinning Nuclear Magnetic Resonance (MAS NMR) in 2D FA DJ perovskites.

Solid-state NMR (ssNMR) spectroscopy allows the elucidation of the local molecular structures of spacer cations and their interactions with A-site and X-site ions within perovskite slabs, and their packing arrangements. In addition, cation dynamics can be studied by relaxation-based or cross-polarization based techniques.[6–9] The $^1$H chemical shifts are sensitive to non-covaletn interactions such as hydrogen bonding or halogen bonding, whereas 207Pb chemical shifts are sensitive to the apical octahedral tilts or Pb-I-Pb bond angles. Specifically, 2D $^1$H-$^1$H exchange spectroscopy (as referred to as spin-diffusion experiments) provide information on through-space proximities between the cage cations and spacer cations. This technique uses different spin-diffusion mixing times synchronized with the rotor period (i.e., integer multiples of sample rotation period), which allow one to probe immediate H-H proximities at sub-nanometer to middle-range proximities of over a nanometer. The on and off-diagonal 2D peaks in such spectra are due to the chemical shifts and through-space proximities, respectively. Intensities of these latter peaks can be adjusted using spin-diffusion mixing time, which allows the spin magnetization to be transferred between the neighboring proton sites in FA and 3AMP cations. A combination of these tools have been used to probe local structures and the cation dynamics in the 2D RP and DJ phases.

## 1.6 Discussion of pair distribution functions (PDFs) calculated for polymorphous cubic FAPbI3 and tetragonal FA DJ n=2.

The pair distribution functions (PDFs) calculated for polymorphous cubic FAPbI$_3$ and tetragonal FA DJ n=2 is showed in Fig. S20. Atomic contributions from the FA or spacer molecules were excluded from our calculations. Compared to the idealized PDFs obtained from the average atomic positions related to the high-symmetry cubic Pm3m or tetragonal P4/mmm crystallographic structures (vertical grey dashed lines), the peaks of the PDFs of the polymorphous structures are shifted and broadened, as shown in Fig. S20 (a, c). These effects arise from a distribution of locally disordered unit cells within the polymorphous networks, as seen before in Ref.[10] This approach allows accounting for the random distribution of atomic positions related to slow relaxational motions (Raman scattering, Fig. 4a) By adding thermal disorder due to lattice vibrations (anharmonic phonons) at 300 K (Fig. S20 (b, d)), the PDFs are further broadened, yielding a better comparison with measured PDFs[11].

## 1.7 Discussion of experimental estimation of exciton diffusion length and diffusivity in n=3 and n=2 DJ FA crystals

Steady state PL diffusion measurements allow extracting exciton diffusion length and diffusivity in 2D materials, by combining PL profiles around the excitation spot and time resolved PL (TRPL) measurements on the same samples.[12] In steady state conditions, the solution of the diffusion equation for the radial distribution of the exciton concentration $n(r)$ corresponds to the convolution between the laser's Gaussian profile and the modified Bessel function of the second kind $K_0$: $n(r) \propto \int_{-\infty}^{+\infty} K_0\left(\frac{r}{L_D}\right) e^{-(r-r')^2/w^2} dr'$, where $L_D$ is the exciton diffusion length and $w$ is the radius of the laser spot. At some distance away from the laser spot $r \gg L_D$, the modified Bessel function can be approximated by $K_0\left(\frac{r}{L_D}\right) \sim \left(\frac{\pi L_D}{2r}\right)^{1/2} e^{-r/L_D}$ and the convolution by the delta-like Gaussian spot leads to $n(r) \propto \left(\frac{L_D}{r}\right)^{1/2} e^{-r/L_D}$. The $L_D$ value of $1.8 \mu m$ ($0.83 \mu m$) was estimated from an asymptotic fit at low excitation power respectively for the n=3 (n=2). The exciton diffusivity $D_X$ can then be extracted by combining the exciton diffusion length $L_D$ with the exciton lifetime $\tau_X$ deduced from TRPL measurements: $L_D = \sqrt{2 D_X \tau_X}$ (notice that the definition of the diffusivity differs by a factor of $\sqrt{2}$ from the one of ref. [12] to match the definition of ref [13] of the main manuscript).[12] The TRPL signal in the case of the n=3 crystal is better described by a sum of multiexponential components. As the exciton diffusion length is estimated at low excitation power and far away from the laser spot, the relevant exciton lifetime $\tau_X$ to estimate the related exciton diffusivity is the long-time TRPL component $\tau_X \sim 3.68 ns$. A high value of the exciton diffusivity is deduced from the present analysis for the n=3 DJ FA compound: $D_X \sim 4.4 cm^2 s^{-1}$.

For the n=2 compound, assuming a similar exciton lifetime, a significantly smaller value of the diffusivity is anticipated $D_X \sim 0.95 cm^2 s^{-1}$.

## 2. Supplementary Figures

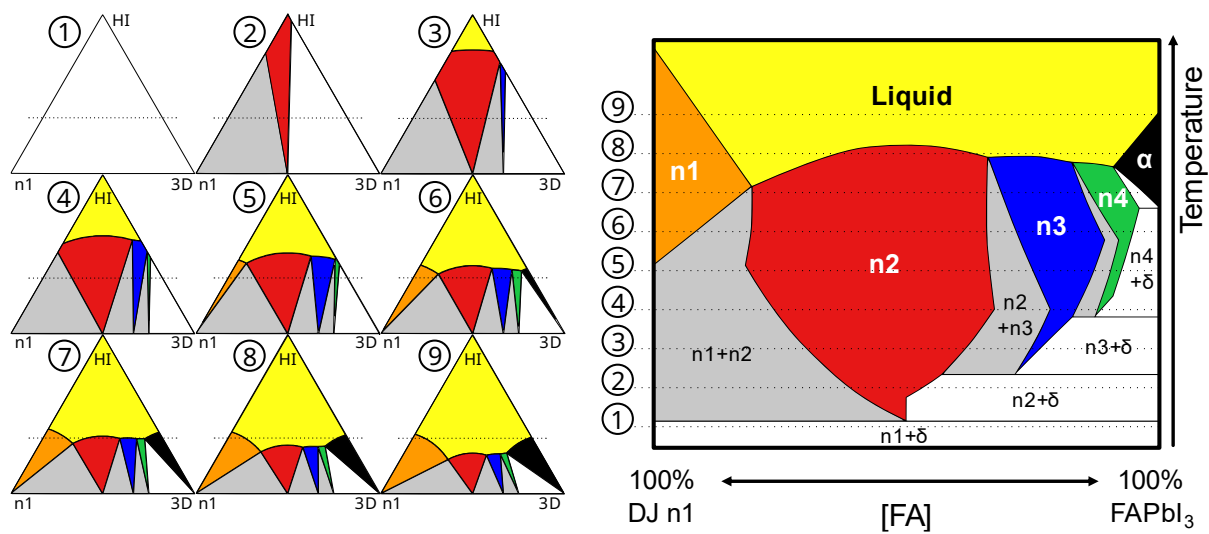

**Fig. S1 | The construction of a binary DJ n1 – FAPbI$_3$ phase diagram from ternary DJn1 – FAPbI$_3$ – HI phase diagrams of increasing temperature.** A specific concentration of crystals in solution was chosen, represented by the dashed horizontal line across each ternary phase diagram. The intersection of each region of the ternary phase diagram with this line was tracked with temperature to build up the binary phase diagram.

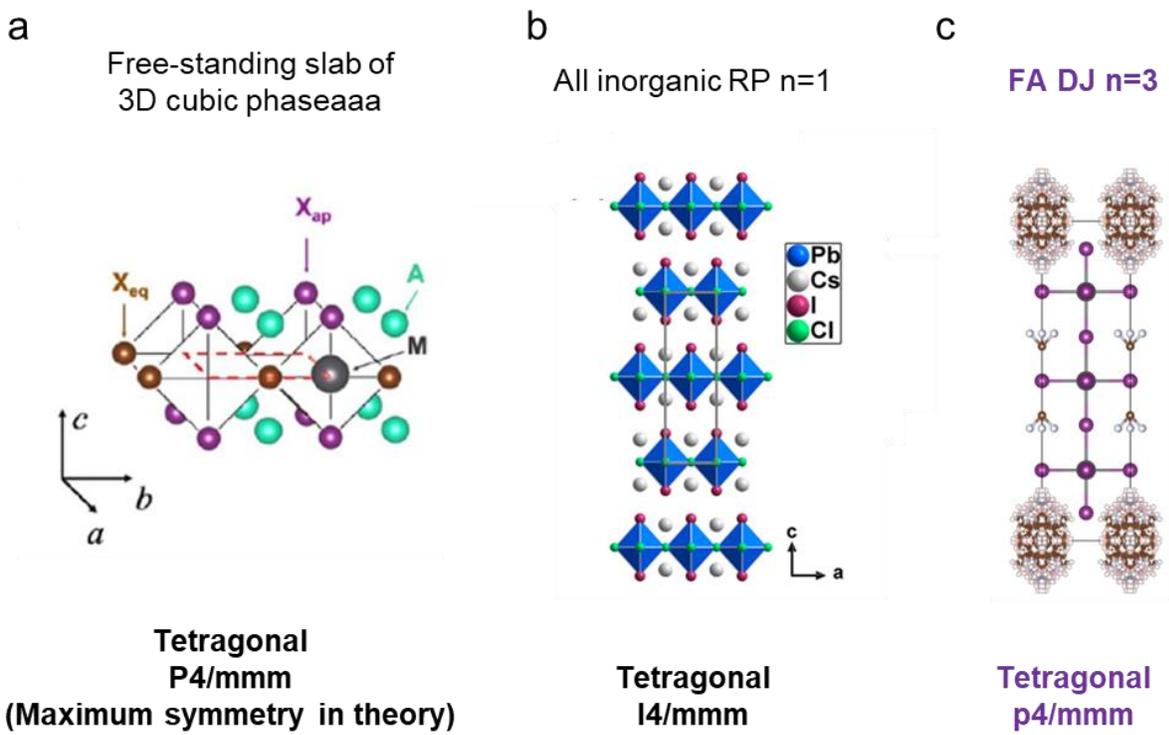

**Fig. S2 | Maximum symmetry in 2D halide perovskite**. (a) A free-standing slab of 3D cubic phase, adapted from ref. [14], licensed under a Creative Commons Attribution (CC BY) license. (b) The all-inorganic RP n=1 perovskites structure. Adapted with permission from ref. [15]. Copyright 2018 American Chemical Society. (c) The FA DJ n=3 structure.

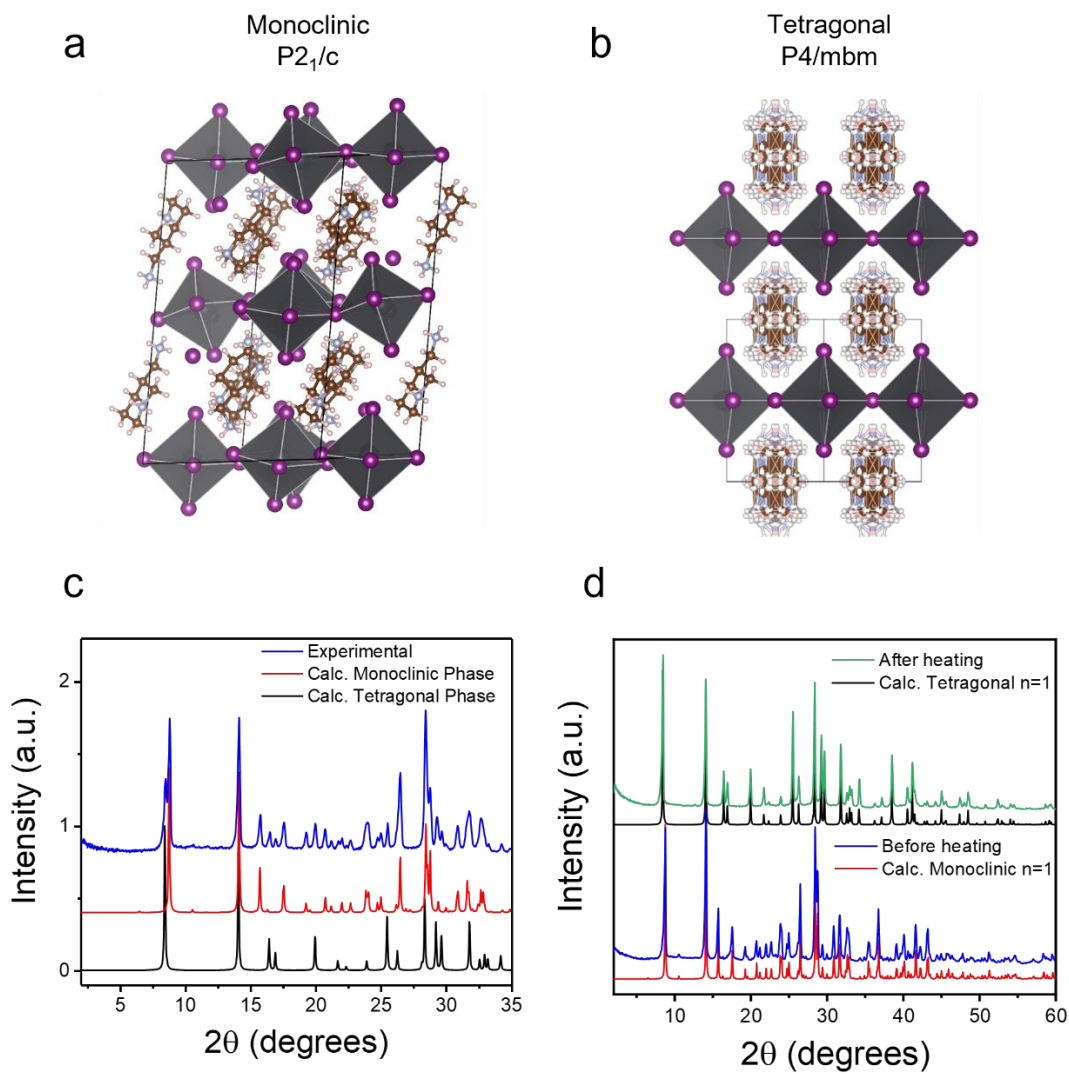

**Fig. S3 | Structure of monoclinic and tetragonal DJ n=1.** (a) The structure of (a) monoclinic and (b) tetragonal 3-(aminomethyl)piperidine (3AMP) n=1 2D perovskite. (c) The experimental powder X-ray diffraction pattern of DJ n=1 containing mixed monoclinic and tetragonal phases and their corresponding calculated patterns. (d) The experimental powder X-ray diffraction patterns of a pure DJ n=1 monoclinic crystal, before and after heating to 190 °C, and the calculated pattern for monoclinic n=1 and tetragonal n=1.

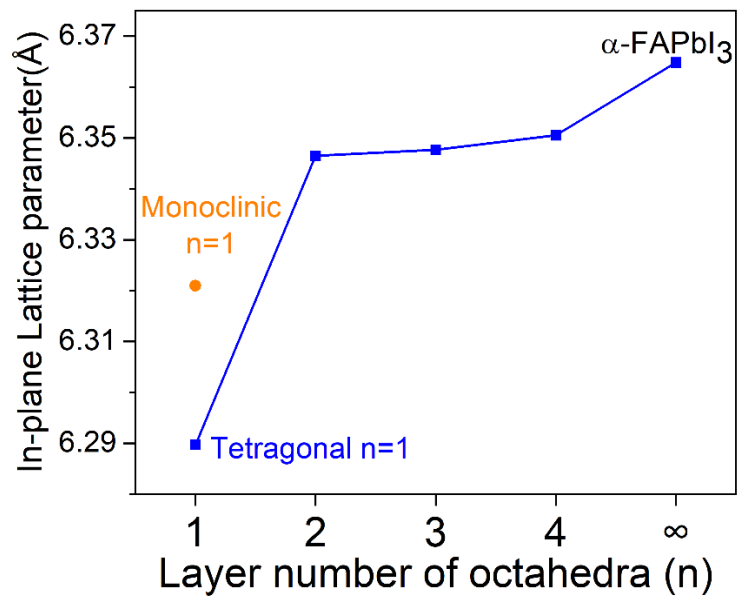

**Fig. S4 | The in-plane lattice parameters of 2D FA DJ perovskites**.

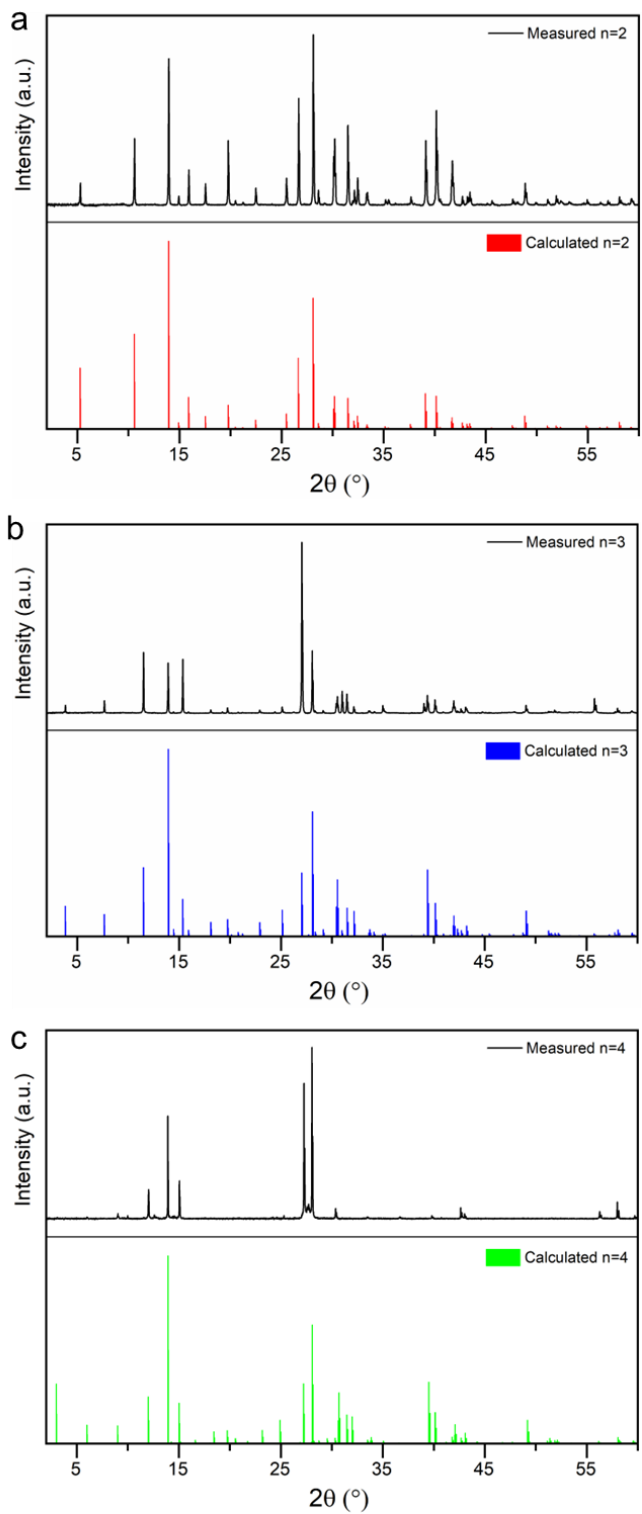

**Fig. S5 | Experimental and calculated powder X-ray pattern of FA DJ (a) n=2, (b) n=3 and (c) n=4**.

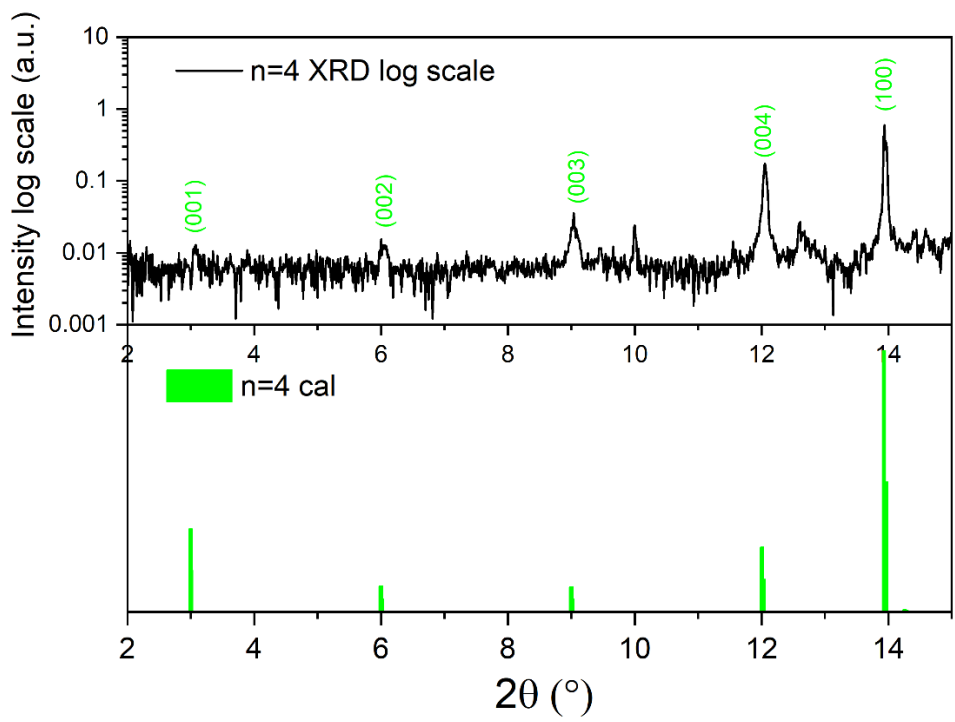

**Fig. S6 | PXRD of FA DJ n=4 at low angle, log-scaled intensity.**

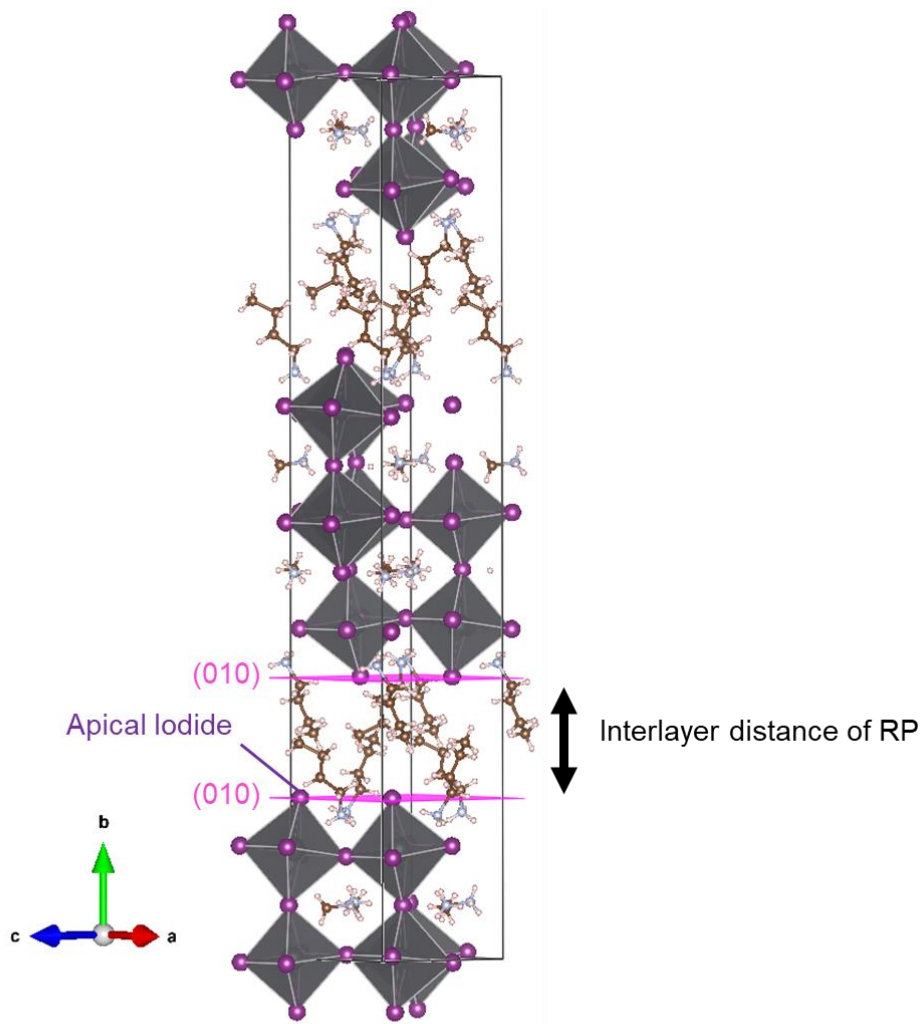

**Fig. S7 | Interlayer distance for RP BA phase.**

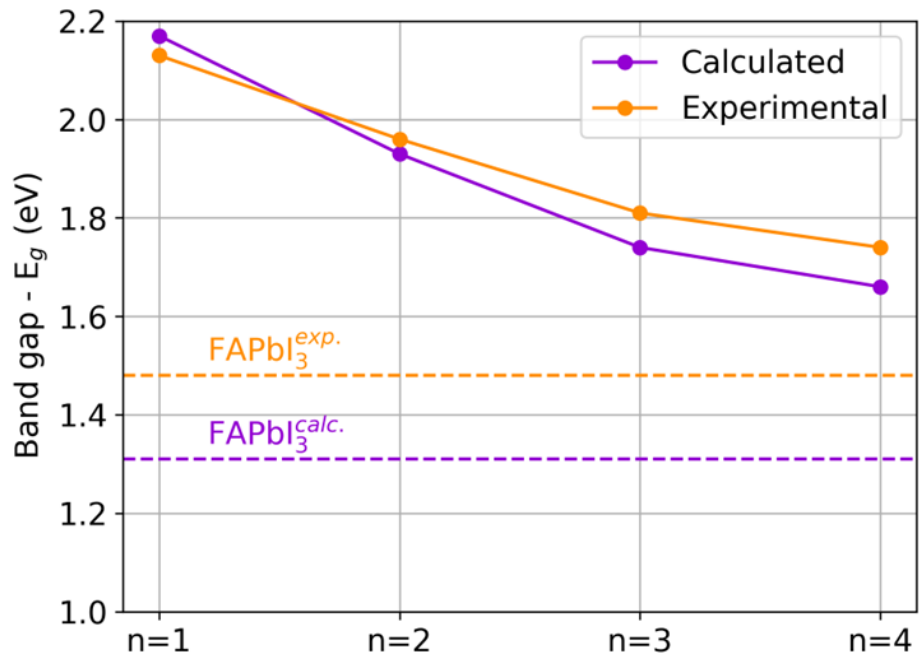

**Fig. S8 | Experimental optical and calculated electronic bandgaps for the FA DJ layered compounds**. We applied a rigid scissor shift of 0.24 eV estimated for $FAPbI_3$ to account for the effects of polymorphism on the 2D FA DJ series.

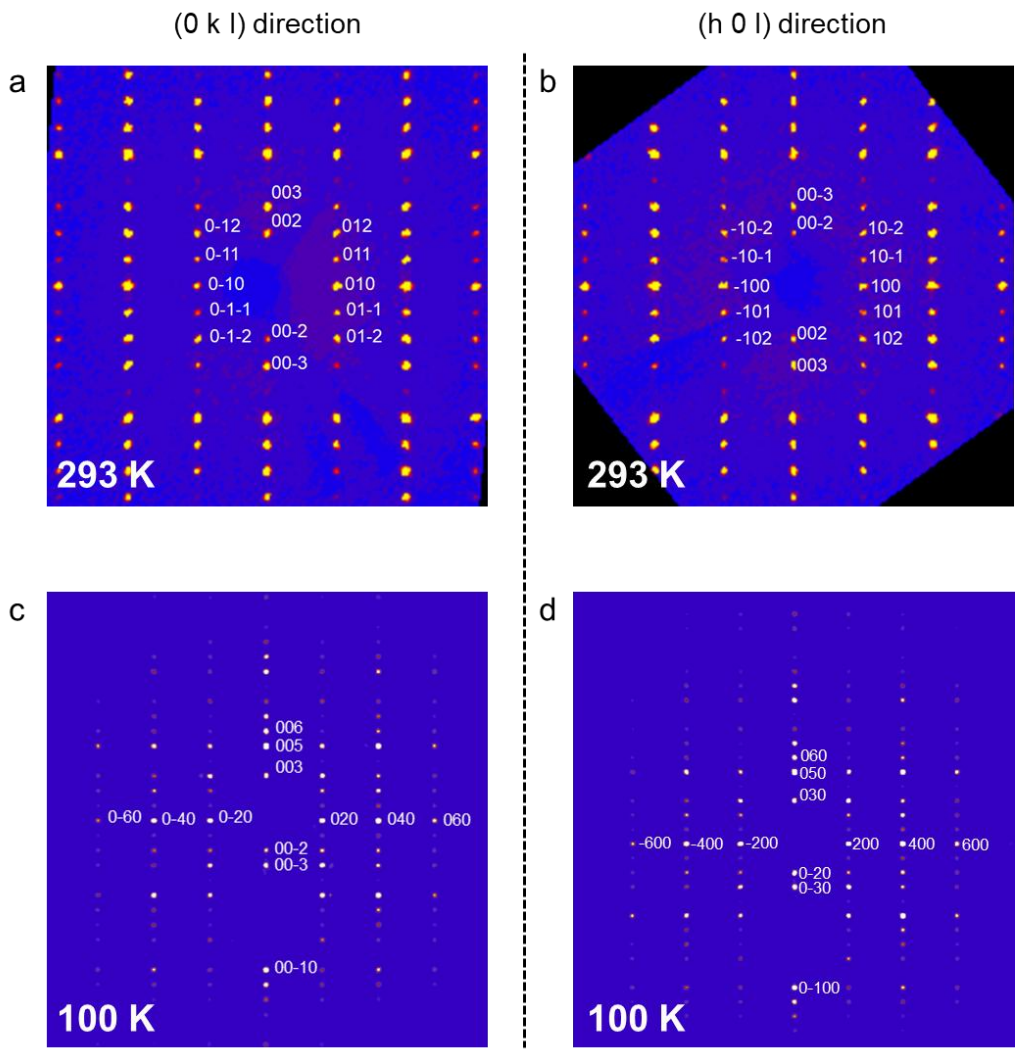

**Fig. S9 | Procession images of FA DJ n=2**, (a) along (0 k l) direction and (b) along (h 0 l) direction at Room temperature (293K), (c) along (0 k l) direction and (d) along (h 0 l) direction at low temperature (100K).

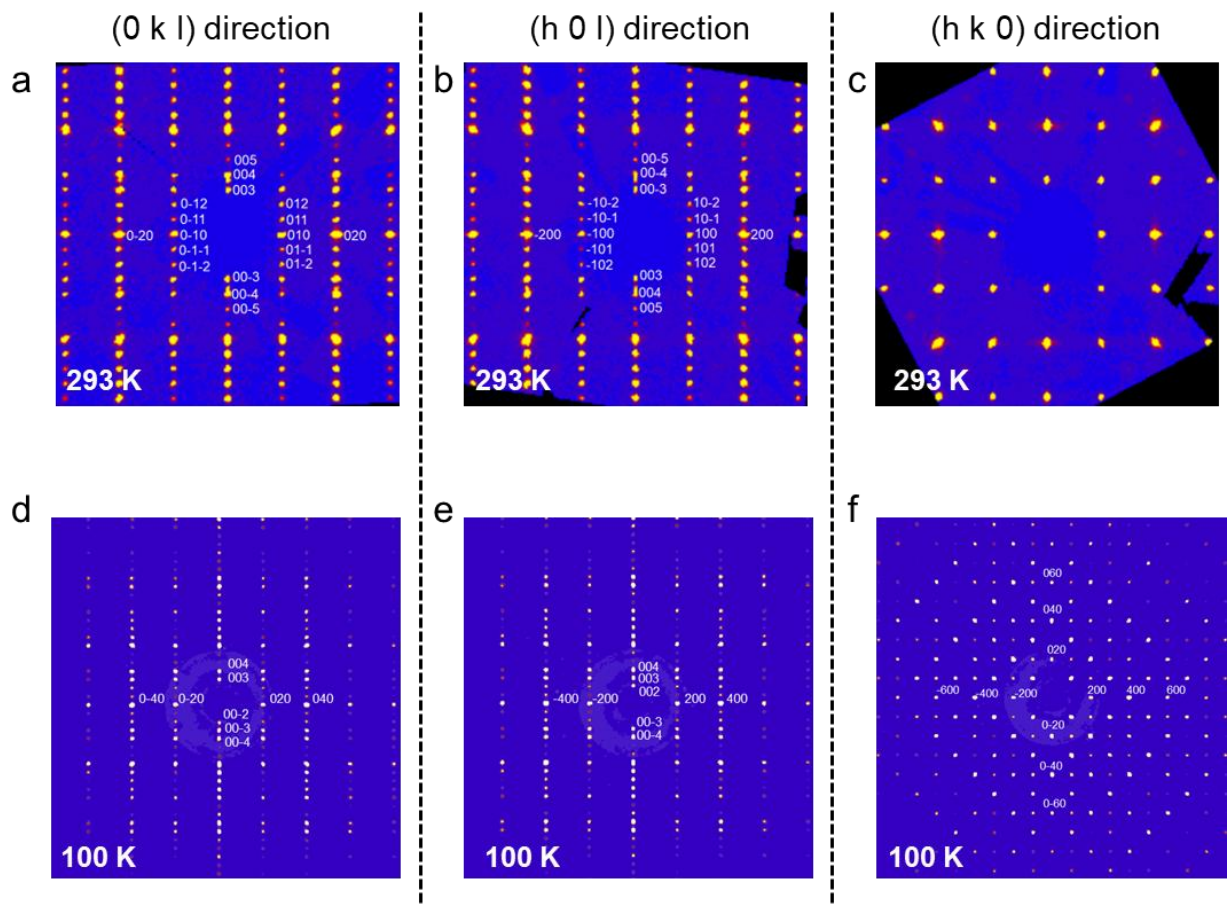

**Fig. S10 | Procession images of FA DJ n=3**, (a) along (0 k l) direction, (b) along (h 0 l) direction, and (c) along (h k 0) direction, at Room temperature (293K), (d) along (0 k l) direction, (e) along (h 0 l ) direction, and (f) along (h k 0) direction, at low temperature (100K).

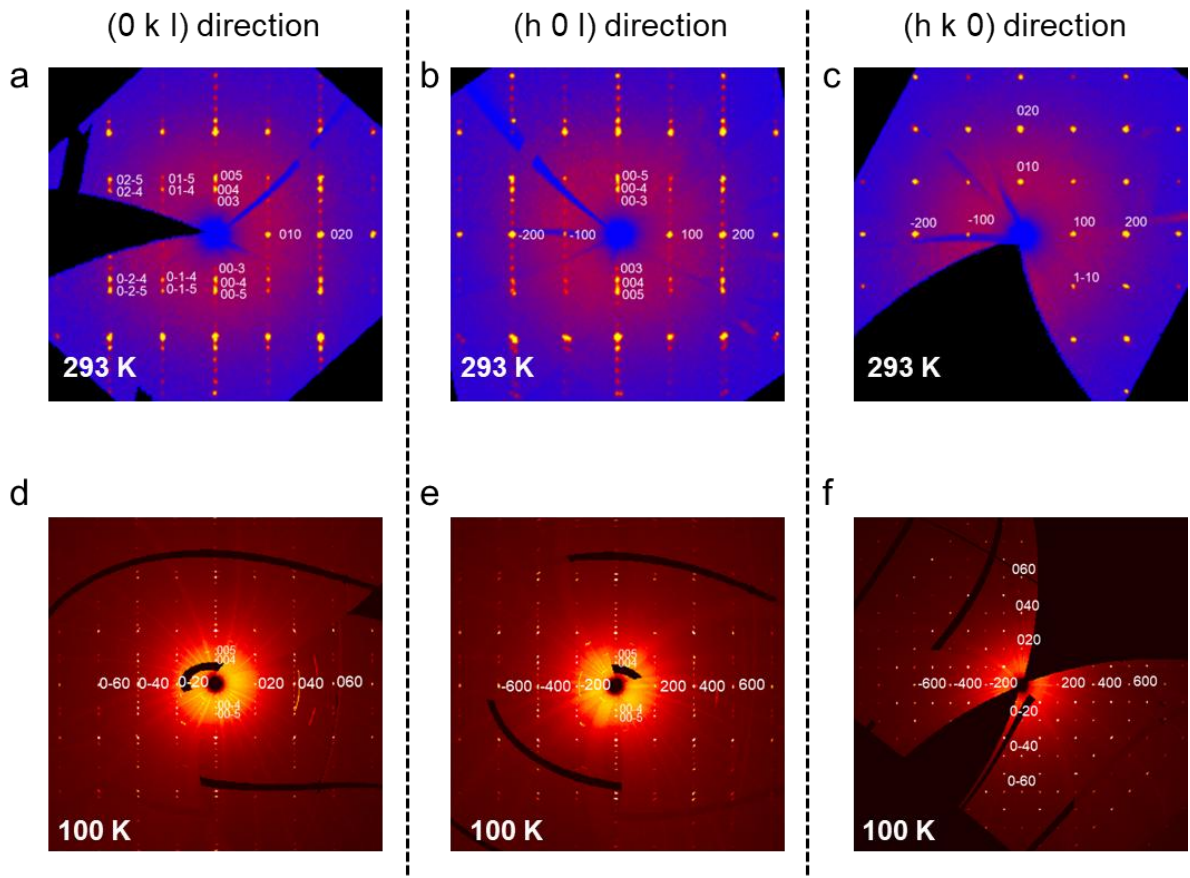

**Fig. S11 | Procession images of FA DJ n=4**, (a) along (0 k l) direction, (b) along (h 0 l) direction, and (c) along (h k 0) direction, at Room temperature (293K), (d) along (0 k l) direction, (e) along (h 0 l) direction, and (f) along (h k 0) direction, at low temperature (100K).

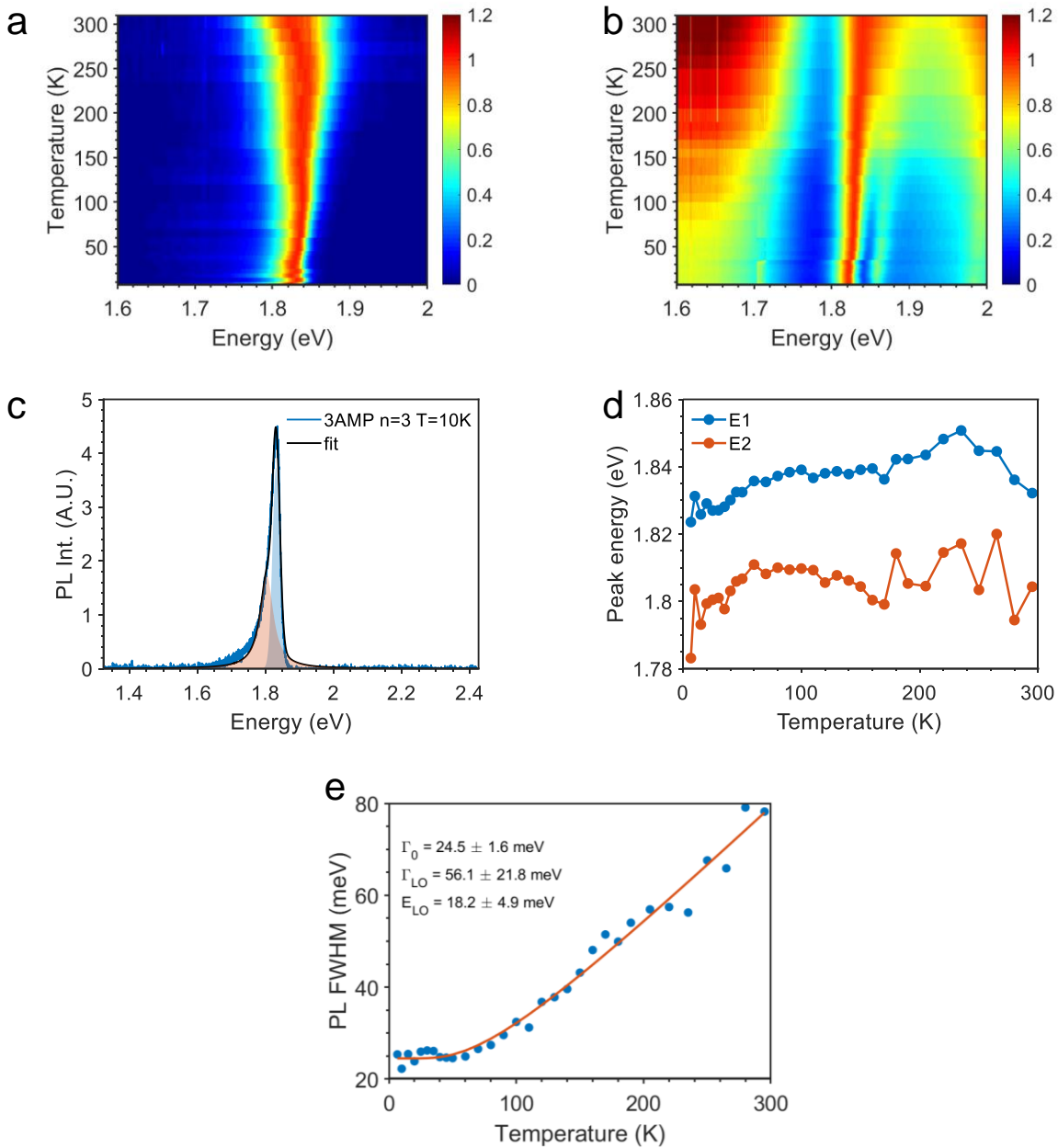

**Fig. S12 | PL broadening analysis of FA DJ n=3.** (a) Normalized steady-state photoluminescence (PL) profile of FA DJ n=3 at temperatures between 6 to 300K. (b) Temperature dependent reflectance spectrum of FA DJ n=3. (c) The raw data and fitting of PL of FA DJ n=3 at 10 K. (d) The extracted peak energy evolution as a function of temperature for free exciton (E1) and bound exciton (E2) states. (e) Temperature-dependent PL FWHM of FA DJ n=3, and the fitting parameters.

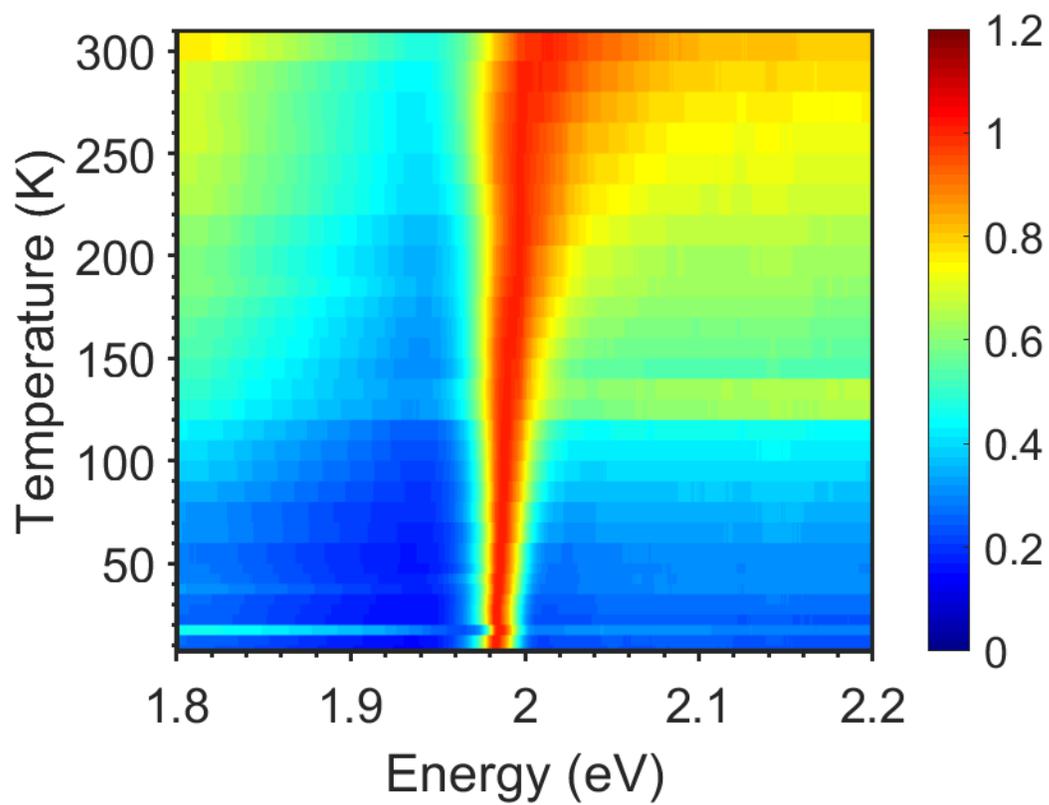

**Fig. S13 | Temperature dependent reflectance spectrum of FA DJ n=2**

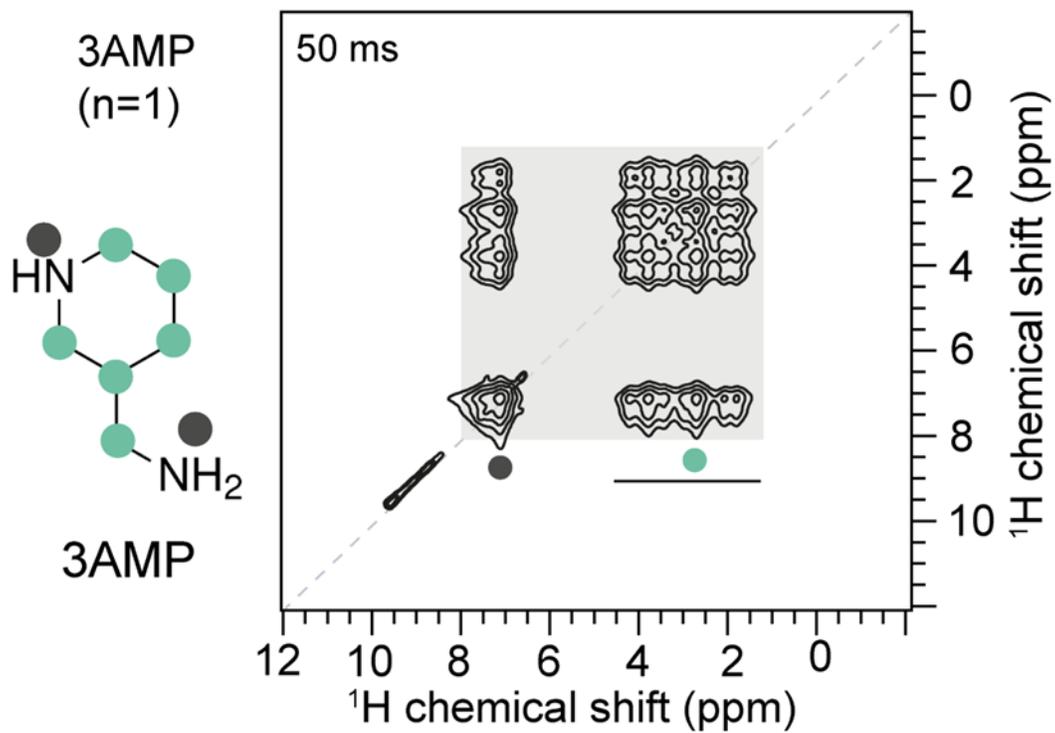

**Fig. S14 | 2D $^1$H-$^1$H spin-diffusion NMR spectra of layered DJ n=1** with a schematic of the 3AMP cation with color dots aiding the spectral analysis.

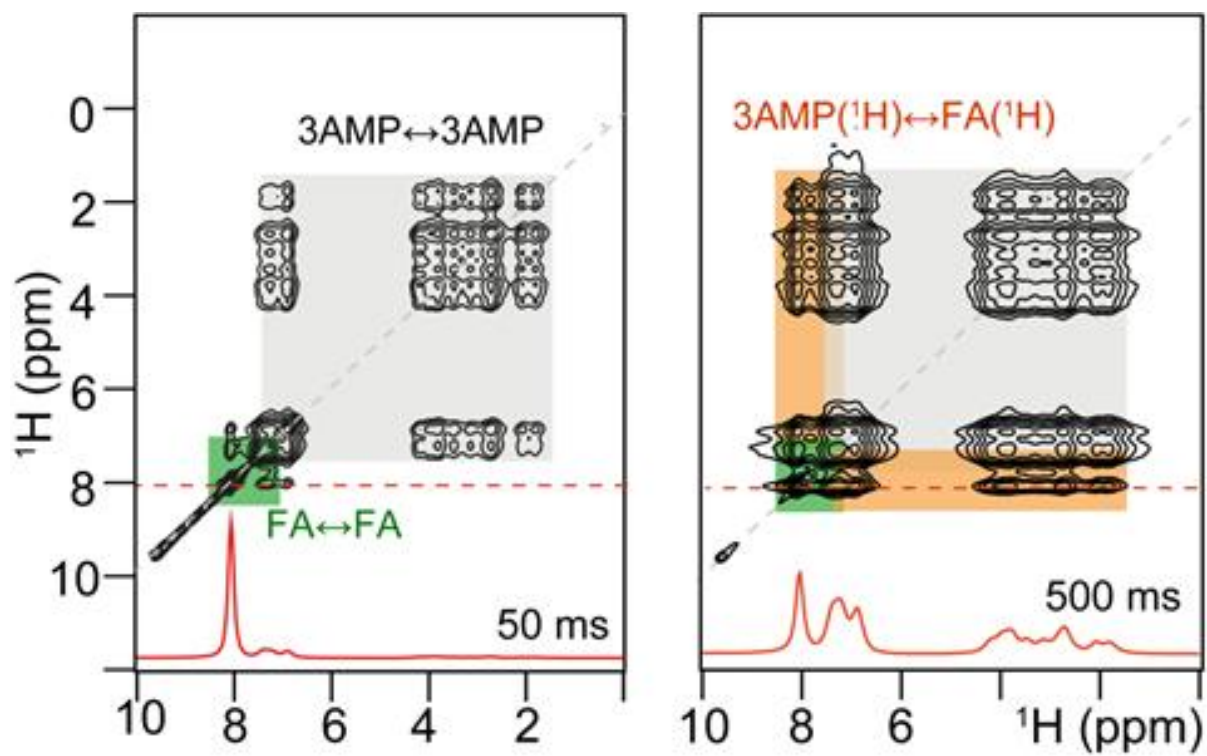

**Fig. S15 | 2D $^1$H-$^1$H spin-diffusion NMR spectra of FA DJ n=2** 2D $^1$H-$^1$H spin-diffusion NMR spectra of FA DJ n=2 acquired at 21.1 T ($^1$H = 900 MHz), room temperature, and 50 kHz MAS using 50 ms (**left**) and 500 ms (**right**) of spin-diffusion mixing time, with the 1D $^1$H NMR spectra plotted on the horizontal projections..

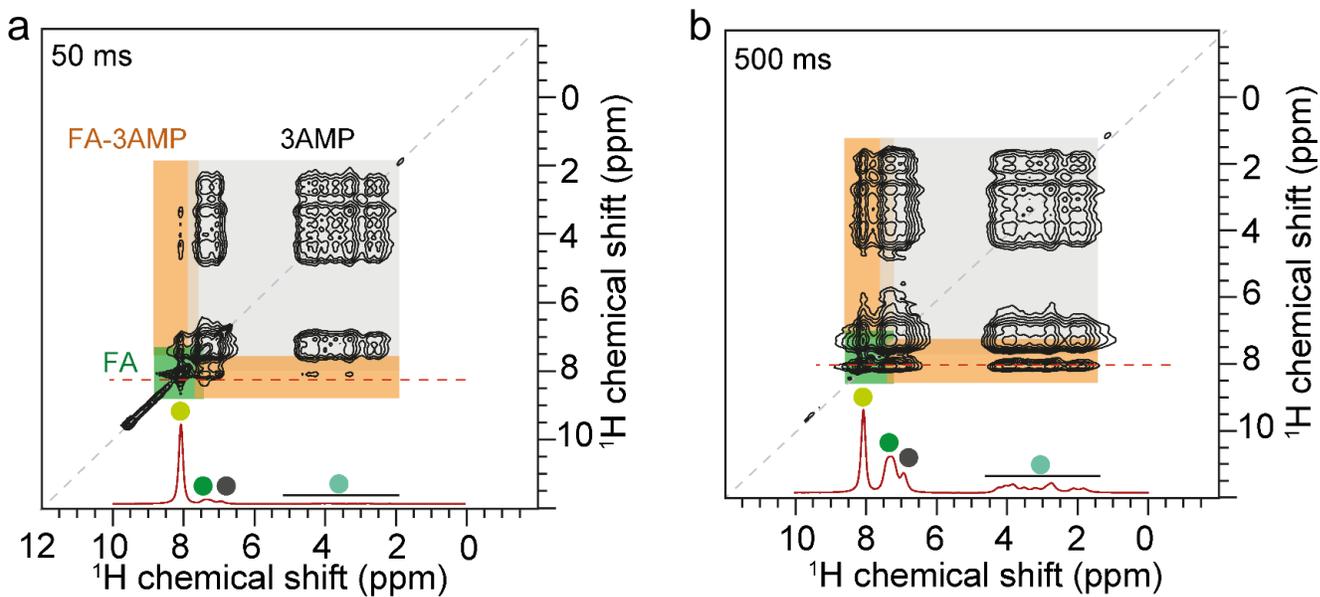

**Fig. S16 | 2D $^1$H-$^1$H spin-diffusion NMR spectra of FA DJ n=3** acquired at 21.1 T ($^1$H = 900 MHz), room temperature, and 50 kHz MAS using 50 ms (**a**) and 500 ms (**b**) of spin-diffusion mixing time, with the 1D $^1$H NMR spectra plotted on the horizontal projections.

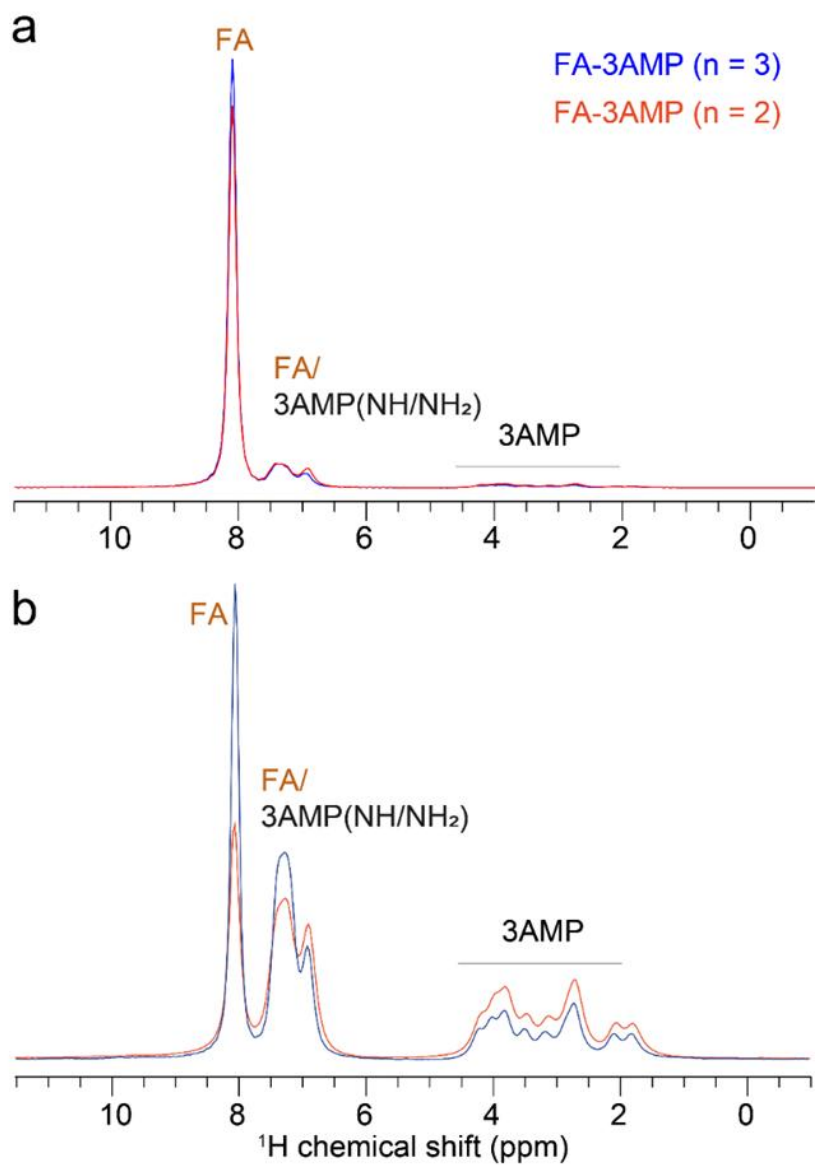

**Fig. S17 | Line-cut row $^1$H spectra obtained of FA DJ**. Line-cut row $^1$H spectra obtained of FA DJ from the 2D $^1$H-$^1$H spin diffusion NMR spectra of FA 3AMP DJ phases obtained with (a) 50 ms and (b) 500 ms at a $^1$H chemical shift of 8.2 ppm (FA, NH). The spectra of n=2 and n=3 phases are depicted in red and blue colors, respectively. All spectra were acquired at 21.1 T ($^1$H = 900 MHz), room temperature, and 50 kHz MAS.

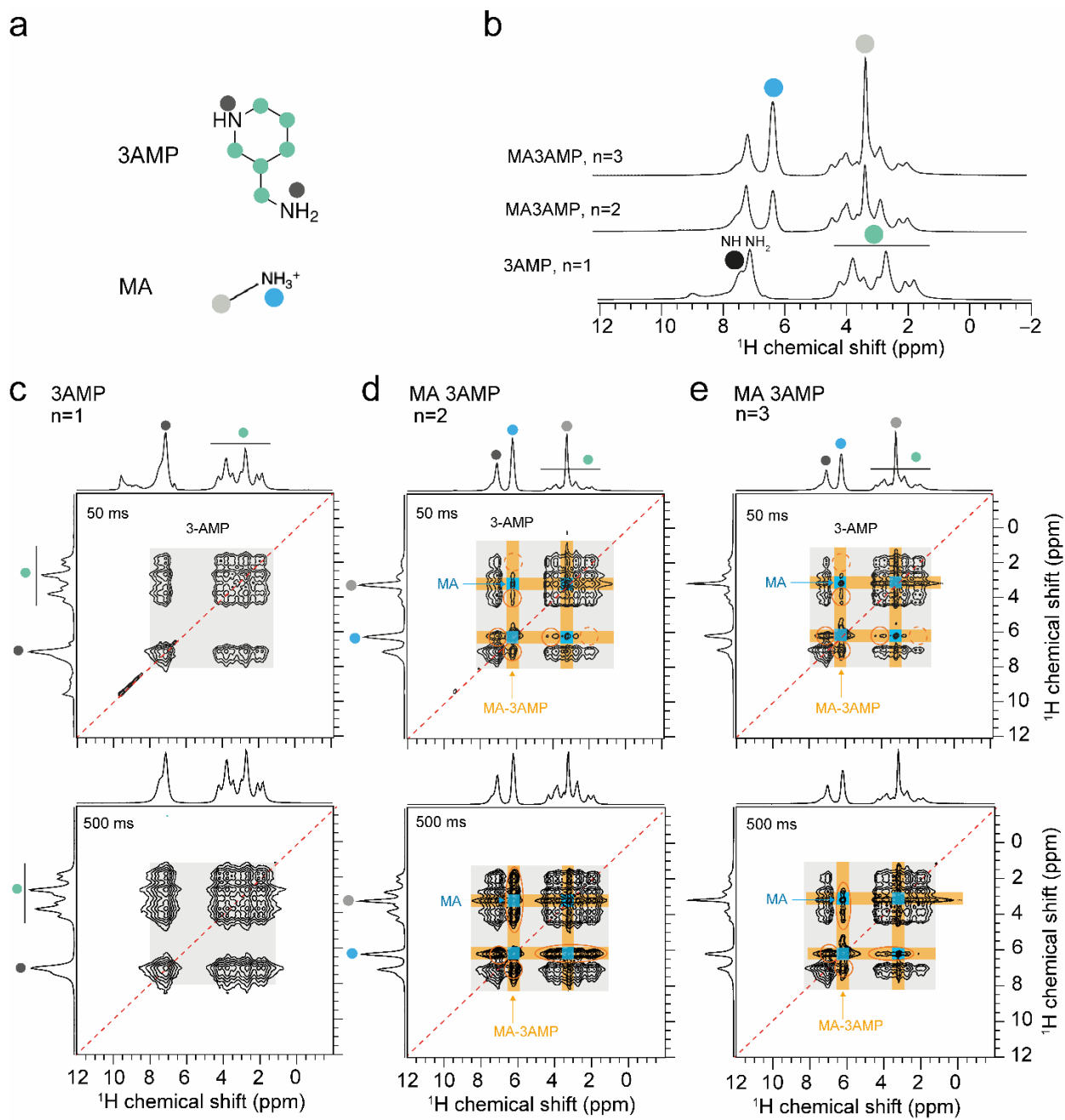

**Fig. S18 | NMR analysis of Methyl ammonium (MA) DJ 2D perovskite**. (a) Schematic of 3AMP and MA cations as color dots to aid spectral interpretation. (b) solid-state 1D 1H NMR spectra of layered DJ phases as indicated. Solid-state 2D 1H-1H SD NMR spectra of layered DJ phases: (c) DJ n=1, (d) MA DJ n=2, and (e) MA DJ n=3 acquired with 50 ms (top) and 500 ms (bottom) of spin diffusion time. Peaks corresponding to the 3AMP spacer and MA cations are depicted in gray and blue boxes. All spectra were acquired at 21.1 T (1H = 900 MHz), room temperature, and 50 kHz MAS.

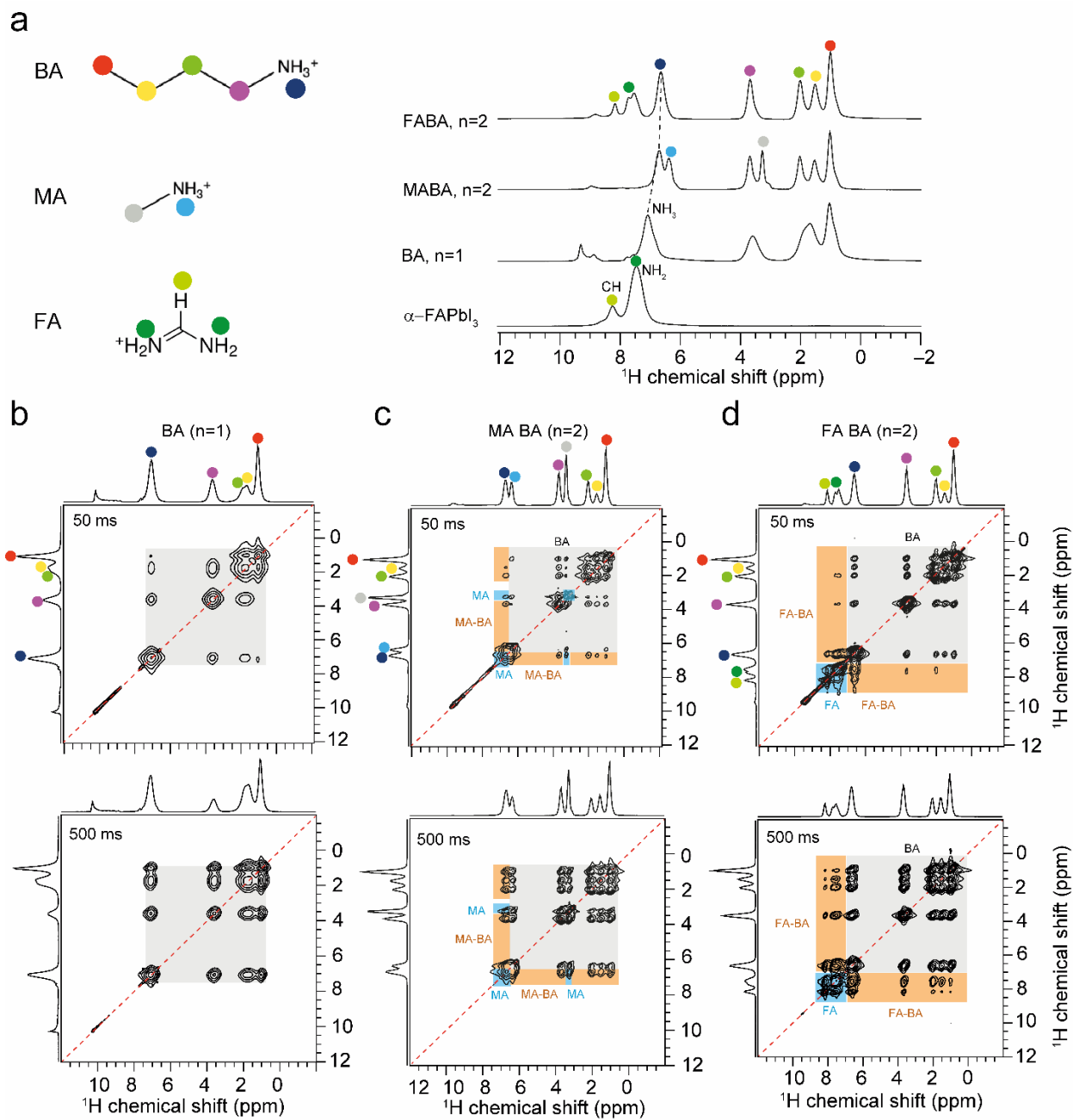

**Fig. S19 | NMR analysis of MA BA and FA BA 2D perovskite**. (a) Schematic of 3AMP and MA cations as color dots. Solid-state (b) 1D NMR spectra of RP phases and 3D perovskites as indicated alongside a schematic of the BA and FA cations with color dots aiding the spectral analysis. 2D 1H-1H spin-diffusion NMR spectra of (c) (BA)PbI$_4$ (n=1), (d) MA BA (n=2), and (e) FA BA (n=2) acquired with 50 ms (top) and 500 ms (bottom) of spin diffusion time, together with the 1D 1H NMR spectra plotted on the horizontal and vertical projections. All spectra were acquired at 21.1 T (1H = 900 MHz), room temperature, and 50 kHz MAS.

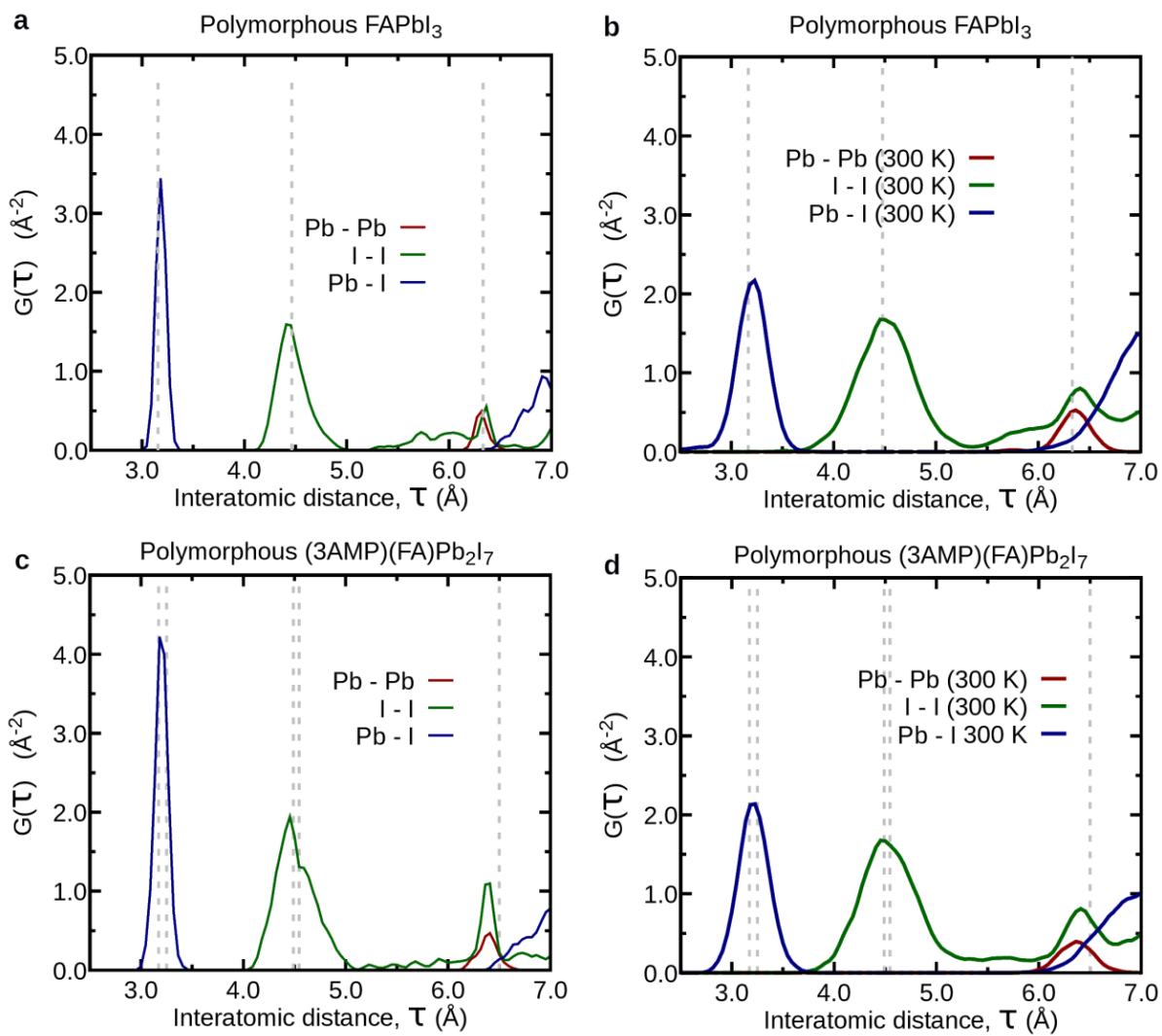

**Fig. S20 | Pair distribution function (PDF) of FA DJ and 3D FAPbI$_3$.** (a,b) Pair distribution function (PDF) of polymorphous cubic FAPbI$_3$ without (a) and with (b) the effect of lattice vibrations at 300 K. Vertical dashed lines represent pair distribution functions of the monomorphous (high-symmetry) FAPbI$_3$. (c,d) As in (a,b) but for FA DJ n=2.

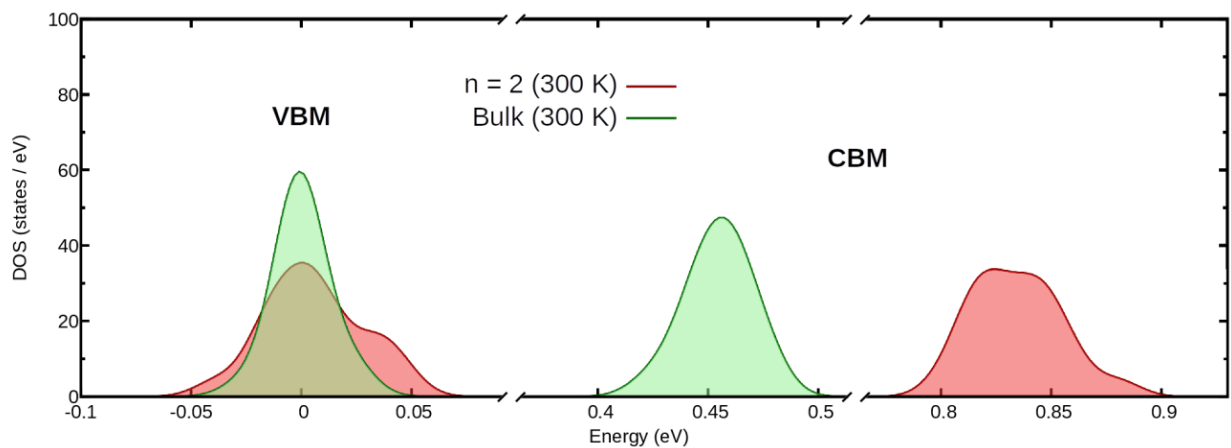

**Fig. S21 | DOS of FA DJ and 3D FAPbI₃**. DFT-PBEsol density of states (DOS) at the valence band maximum (VBM) and conduction band minimum (CBM) of polymorpous FA DJ n=2 (red) and $FAPbI_3$ (green). Calculations include the effect of electron-phonon coupling at 300 K.

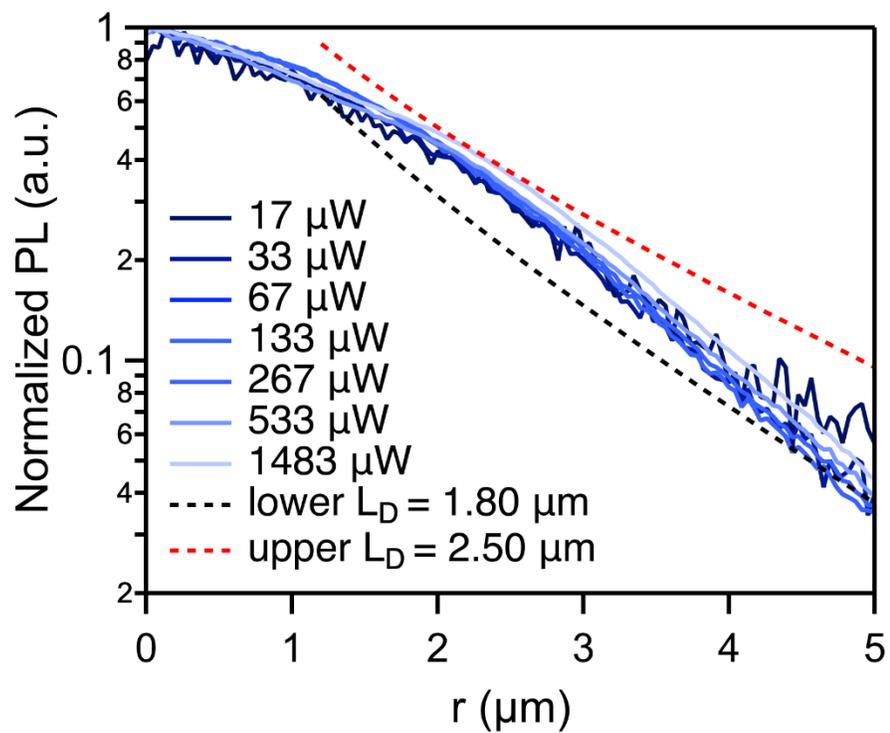

**Fig. S22 | Diffusion lengths derived from fitting of normalized spatial PL profiles of FA DJ n=3**

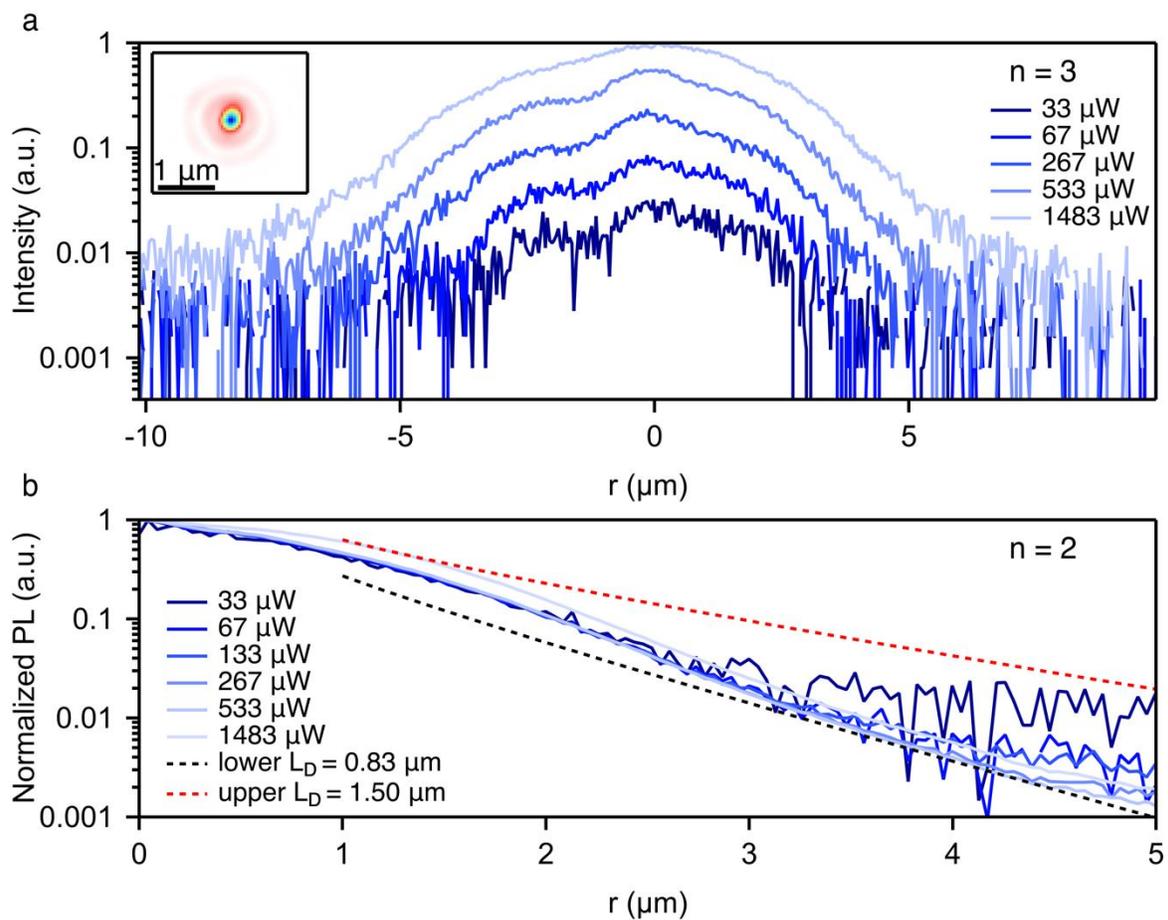

**Fig. S23 | FA DJ diffusion length analysis.** (a) Power dependent diffusion profiles cut from FA DJ n=3 real-space PL images. Inset shows reflection of 532 nm laser with ~ 0.6 µm spot size. (b) Normalized PL intensities for FA DJ n=2 where upper (red) and lower (black) bound diffusion length fittings are 1.5 µm and 0.83 µm respectively.

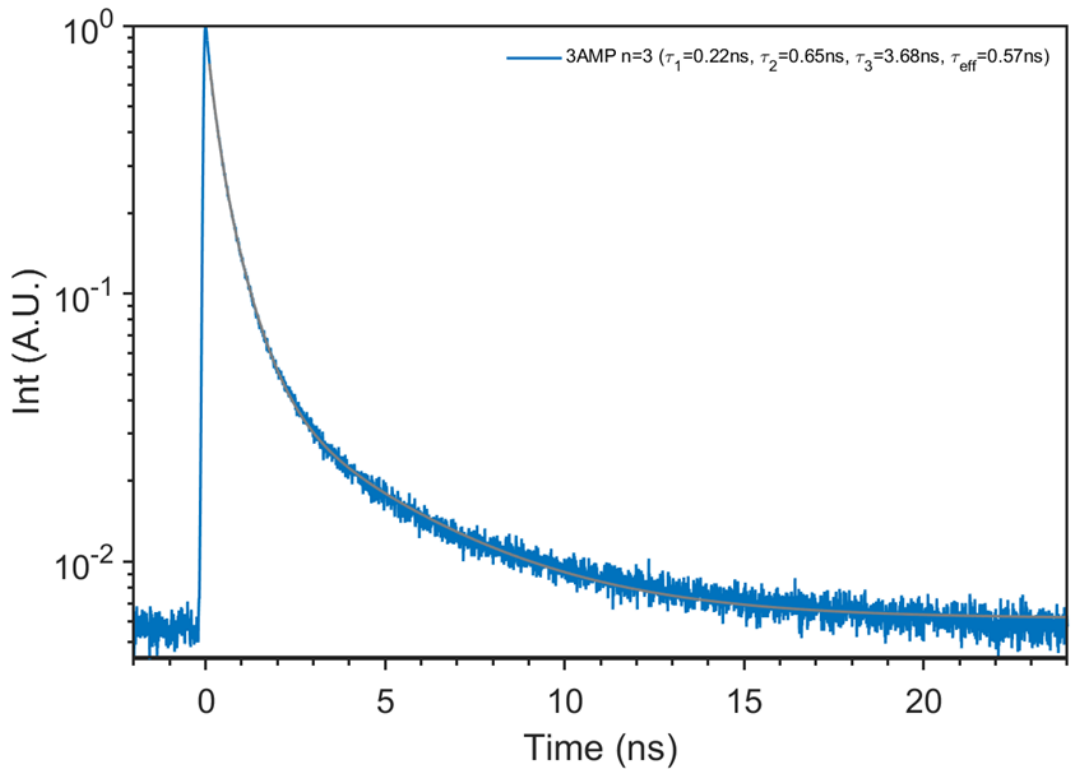

**Fig. S24 | Time-resolved photoluminescence spectroscopy (TRPL) of FA DJ n=3 crystal flake.** The PL was spectrally selected at exciton emission at 680nm, showing non-exponential decay traces which are fitted with triple exponential decay curves ($\tau_1 = 0.22$ ns, $\tau_2 = 0.65$ ns, $\tau_3 = 3.68$ ns) with average lifetime $\tau = 0.57$ ns.

# 3. Supplementary Table
## Table S1 | Crystal Data and Structure Refinement for (3AMP)(FA)$_{n-1}$Pb$_n$I$_{3n+1}$, n=1 to n=4.

| | Compound | | | |
|---|---|---|---|---|
| | (3AMP)PbI$_4$ | (3AMP)(FA)Pb$_2$I$_7$ | (3AMP)(FA)$_2$Pb$_3$I$_{10}$ | (3AMP)(FA)$_3$Pb$_4$I$_{13}$ |
| Empirical formula | C6 H16 I4 N2 Pb | C7 H16 I7 N4 Pb2 | C8 H16 I10 N6 Pb3 | C9 H16 I13 N8 Pb4 |
| Formula weight | 831.00 | 1458.92 | 2086.84 | 2714.76 |
| Temperature | 285(14) K | 293 K | 293 K | 293 K |
| Wavelength | 0.71073 Å | 0.56083 Å | 0.56083 Å | 0.56083 Å |
| Crystal system | Tetragonal | Tetragonal | Tetragonal | Tetragonal |
| Space group | P4/mbm | P4/mmm | P4/mmm | P4/mmm |
| Unit cell dimensions | a = 8.8950(2) Å, α = 90°<br>b = 8.8950(2) Å, β = 90°<br>c = 10.4829(4) Å, γ = 90° | a = 6.3465(9) Å, α = 90°<br>b = 6.3465(9) Å, β = 90°<br>c = 16.722(3) Å, γ = 90° | a = 6.3477(9) Å, α = 90°<br>b = 6.3477(9) Å, β = 90°<br>c = 23.054(5) Å, γ = 90° | a = 6.3505(9) Å, α = 90°<br>b = 6.3505(9) Å, β = 90°<br>c = 29.454(6) Å, γ = 90° |
| Volume | 829.42(5) Å$^3$ | 673.5(2) Å$^3$ | 928.9(3) Å$^3$ | 1187.9(4) Å$^3$ |
| Z | 2 | 1 | 1 | 1 |
| Density (calculated) | 3.327 g/cm$^3$ | 3.597 g/cm$^3$ | 3.730 g/cm$^3$ | 3.795 g/cm$^3$ |
| Absorption coefficient | 17.588 mm$^{-1}$ | 10.946 mm$^{-1}$ | 11.686 mm$^{-1}$ | 12.070 mm$^{-1}$ |
| F(000) | 720 | 621 | 882 | 1143 |
| θ range for data collection | 3.239 to 29.988° | 1.922 to 24.992° | 2.091 to 24.987° | 2.182 to 19.981° |
| Index ranges | -12<=h<=12, -12<=k<=10, -14<=l<=14 | -9<=h<=9, -9<=k<=9, -25<=l<=24 | -9<=h<=9, -9<=k<=9, -34<=l<=34 | -7<=h<=7, -7<=k<=7, -35<=l<=35 |
| Reflections collected | 14588 | 15960 | 29086 | 9528 |
| Independent reflections | 694 [R$_{int}$ = 0.0232] | 796 [R$_{int}$ = 0.0677] | 1086 [R$_{int}$ = 0.0406] | 773 [R$_{int}$ = 0.1328] |
| Completeness to θ = 25.242° | 99.8% | 99.8% | 98.6% | 99.3% |
| Refinement method | Full-matrix least-squares on F$^2$ | Full-matrix least-squares on F$^2$ | Full-matrix least-squares on F$^2$ | Full-matrix least-squares on F$^2$ |
| Data / restraints / parameters | 694 / 65 / 82 | 796 / 65 / 91 | 1086 / 71 / 100 | 773 / 26 / 63 |
| Goodness-of-fit | 1.056 | 0.829 | 1.068 | 0.984 |
| Final R indices [I > 2σ(I)] | R$_{obs}$ = 0.0276, wR$_{obs}$ = 0.0709 | R$_{obs}$ = 0.0290, wR$_{obs}$ = 0.0641 | R$_{obs}$ = 0.0285, wR$_{obs}$ = 0.0799 | R$_{obs}$ = 0.0637, wR$_{obs}$ = 0.1496 |
| R indices [all data] | R$_{all}$ = 0.0348, wR$_{all}$ = 0.0758 | R$_{all}$ = 0.0672, wR$_{all}$ = 0.0684 | R$_{all}$ = 0.0434, wR$_{all}$ = 0.0837 | R$_{all}$ = 0.1395, wR$_{all}$ = 0.1964 |
| Largest diff. peak and hole | 1.587 and -1.861 e·Å$^{-3}$ | 0.903 and -1.627 e·Å$^{-3}$ | 1.238 and -1.954 e·Å$^{-3}$ | 6.944 and -2.056 e·Å$^{-3}$ |

**Table S2 | Effective lattice parameter.** a is the real lattice parameter whereas L correspond to the effective parameter , which is essentially the distance between two in-plane neighboring Pb atoms.

| | |
|---|---|
| 3D α-phase FAPbI$_3$, cubic Pm-3m | a=L |
| FA DJ n=2 to n=4, tetragonal, P4/mmm | a=L |
| DJ n=1, tetragonal, P4/mbm | $\frac{a}{\sqrt{2}} = L$ |
| DJ n=1, monoclinic, P2$_1$/c | $\sqrt{ab}/2 = L$ |

**Table S3 | Extracted PL linewidth and fitting parameters of FA DJ perovskites.**
The extracted values for inhomogeneous exciton linewidth ($\Gamma_0$), strength of LO phonon coupling ($\Gamma_{LO}$), and LO phonon energy ($E_{LO}$) are listed for n = 2 and n = 3 perovskites, along with those reported for 3D FAPbI$_3$ films and nanocrystals.[5,16] The parameters are fitted using the formula of section 1.4.

| Sample | $\Gamma_0$ (meV) | $\Gamma_{LO}$ (meV) | $E_{LO}$ (meV) |
|---|---|---|---|
| n=2 | 29.6 ± 0.5 | 20 ± 4 | 8.5 ± 1.2 |
| n=3 | 24.5 ± 1.6 | 56 ± 22 | 18.2 ± 4.9 |
| FAPbI$_3$ | 19 ± 1 | 40 ± 5 | 11.5 ± 1.2 |
| FAPbI$_3$ (NC) | 1.5 | 27 | 10.7 |

**Table S4 | Optical dielectric constant (ε∞) and electronic band gaps (Eg).** Optical dielectric constant (ε∞) calculated within DFPT, and electronic band gaps ($E_g$) using the DSH hybrid functional, for the FA DJ series and for FAPbI$_3$, with (1.31) and without (1.07) the correction due to polymorphism.

|  | ε∞ (PBE) | $E_g$ (DSH) |
|---|---|---|
| **n=1** | 5.18 | 1.93 |
| **n=2** | 5.76 | 1.69 |
| **n=3** | 6.04 | 1.50 |
| **n=4** | 6.20 | 1.42 |
| **FAPbI3** | 6.97 | 1.07 |
| **poly-FAPbI3** | 6.87 | 1.31 |

# 4. Methods
## 4.1 Crystal synthesis

**Reagents**: PbO (99.9%), hydroiodic acid (HI, 57 wt % in H2O, distilled, stabilized, 99.95%), hypophosphorous acid solution ($H_3PO_2$, 50 wt % in H2O) were purchased from Sigma-Aldrich. 3-(aminomethyl)piperidine (3AMP, 98%) was purchased from TCI, formamidinium chloride (FACl) was purchased from Great cell solar. All chemicals were used as received.

**Synthesis of $(3AMP)(FA)_{n-1}Pb_nI_{3n+1}$, n=1 (Large batch powder crystal):** For n=1, PbO powder (223.2 mg, 1mmol) was dissolved in a mixture of 57% w/w aqueous HI solution (5.0 mL, 39.2 mmol) at room temperature (25 °C) under constant magnetic stirring for 10 mins, which formed a bright yellow solution ($PbI_2$). In a separate beaker, 3AMP (101.6 μL, 0.85 mmol) was neutralized with $H_3PO_2$ 50 wt % in aqueous (1.0 mL, 9.1 mmol) in an ice bath resulting in a clear colorless solution. Then the 3AMP solution was added into the $PbI_2$ solution under heating at 230 °C (hotplate temperature) with constant stirring. The addition initially produced red precipitates, which were slowly dissolved under heating. Then the solution was cooled down to room temperature and red needle-like crystals precipitated. The product n=1 sometime contains both monoclinic phase and tetragonal phase. If the crystal is heated up to 190 °C, it will be locked at tetragonal phase even after being cooled down to room temperature (Fig. S1).

**Synthesis of $(3AMP)(FA)_{n-1}Pb_nI_{3n+1}$, n=2-3 (Large batch powder crystal):** Since the FACl is highly hygroscopic, it was stored and weighed in the glovebox under inert conditions.

For n=2, firstly, FACl (483mg, 6 mmol) was added in to a 50 ml glass conical flask sealed with a glass stopper prior to removal from the glovebox. The subsequent steps were conducted on a hot plate with vigorous stirring inside a fume hood. The glass stopper was removed and 57% wt % aqueous HI solution (20.0 mL, 152 mmol) and $H_3PO_2$ 50 wt % aqueous (2.0 mL, 18.2 mmol) were quickly added to dissolve the FACl until a clear yellow solution was obtained. PbO powder (2678.4 mg, 12 mmol) was added into the FACl solution to form a yellow suspension at room temperature, then the hotplate was set to 230 °C (500 RPM stirring), until the color of suspension turned black. In a second beaker, 3AMP (427.4 μL, 3.6 mmol) was neutralized with 50 wt % aqueous $H_3PO_2$ (3.0 mL, 27.3 mmol) in an ice bath resulting in a clear colorless solution. Then the 3AMP/ $H_3PO_2$ solution was added into the black suspension in the first flask with vigorous stirring, within 1~2 minutes a clear yellow solution was obtained and it was further stirred for 5 mins. From this point, the temperature of hotplate was turned down to 120 ˚C (cooling rate of our ceramic hotplate is around 1 ˚C/4s). After the temperature reached 120 C, the solution was left at this temperature for 1 hour, and dark-red plate-like crystal slowly precipitated out. The crystal was then isolated rapidly without further cooling by suction filtration, followed by drying on the filtration funnel for a further 5 min. Finally, the crystal was put in a clean vial and dried in vacuum at 60 °C overnight.

For n=3, firstly, FACl (655mg, 8.1 mmol) was added in to a 50 ml glass conical flask sealed with a glass stopper prior to removal from the glovebox. The subsequent steps were conducted on a hot plate with vigorous stirring inside a fume hood. The glass stopper was removed, $H_3PO_2$ 50 wt % aqueous (2.0 mL, 18.2 mmol) and 57% wt % aqueous HI solution (18.0 mL, 136.8 mmol) were quickly added to dissolve the FACl until a clear yellow solution was obtained. PbO powder (2566.8 mg, 11.5 mmol) was added into the FACl solution to form a yellow suspension at room temperature, then the hotplate was set to 230 °C (500 RPM stirring), until the color of suspension turned black. In a second beaker, 3AMP (150 μL, 1.3 mmol) was neutralized with 50 wt % aqueous $H_3PO_2$ (4.0 mL, 36.4 mmol) in an ice bath resulting in a clear colorless solution. Then the 3AMP/ $H_3PO_2$ solution was added into the black suspension in the first flask with vigorous stirring, within 1~2 minutes a clear yellow solution was obtained and it was further stirred for 5 mins. From this point, the temperature of hotplate was turned down to 120 ˚C (cooling rate of our ceramic hotplate is around 1 ˚C/4s). After the temperature reaches 120 ˚C, the solution is left at this temperature for 1 hour, and black plate-like crystal slowly precipitated out. The crystal was then isolated rapidly without further cooling by suction filtration, followed by drying on the filtration funnel for a further 5 min. Finally, the crystal was put in a clean vial and dried in vacuum at 60 °C overnight.

To avoid the corrosion of the vacuum oven (mostly made of steel), crystal can be washed using heptane to get rid of the residue HI before drying in the oven. (Certainly, this washing is optional.)

Synthesis of n=1, n=2 and n=3 powders are highly reproducible and scalable. For powder synthesis of n=4, we found it is hard to obtain pure product as the yellow, δ-phase $FAPbI_3$ always form together as the impurity. We were only able to synthesize pure n=4 using a previously reported KCSC method, which is introduced below. However, it is small single crystal (up to mm size or smaller) and not scalable. More discussion on the synthesis can be found in SI section 1.3.

**Synthesis of $(3AMP)(FA)_{n-1}Pb_nI_{3n+1}$, n=4 (thin large crystal):** The n=4 large thin crystal was synthesized using our previous reported kinetically controlled space confinement (KCSC) method.[4] First, solution of n=3 was prepared using the stoichiometry introduced above. After the addition of 3AMP/ $H_3PO_2$ solution was added into the black suspension in the first flask, yellow solution was obtained. From this point, instead of cooling it down, 1ml of boiling yellow solution is taken out and added in to 500 μL HI solution (57% wt % aqueous) in a separate vial.

   Glass was used as the substrate for the 2D perovskite growth. Glass substrates were cut into 1-inch* 1-inch squares, cleaned in soap water, acetone, isopropanol by ultrasonication for 20 min each; then dried by argon. The substrates were transferred into a UV-Ozone cleaner and cleaned for 10 mins. The substrates were put on a hot plate, 12.5 μL of the parent solution was dropped onto the glass surface, another glass slide was put on top to fully cover the bottom glass and annealed at 80 °C for 7 hours. Then the top glass is removed, the crystal with the bottom glass is placed on spin coater, heptane as the washing solvent was dropped (80 μl) instantly on the crystal while spin coated at 3000 r.p.m for 30 seconds to remove all the residue parent acid solution.

The synthesis n=4 is not as reproducible as the lower n, sometimes multiple attempts are necessary. Yellow needle-like crystal, which is $NH_4PbI_3·2H_2O$, could be observed[17] occasionally as impurity if the reaction period is too long, as the partial decomposition of formamidine to ammonia in acidic media.[18–20]

This method for synthesis of large thin crystal is applicable to n=1 to n=3. Each of them uses the solution with corresponding stoichiometry. Specifically, for n=1 large thin crystal, solution of n=1 was prepared using the stoichiometry introduced above, and then annealed at 70 °C for 5 hours; For n=2 crystal, solution of n=2 was prepared using the stoichiometry introduced above, and then annealed at 70 °C for 4 hours; For n=3 crystal, solution of n=3 was prepared using the stoichiometry introduced above, and then annealed at 70 °C for 4 hours.

## 4.2 Construction of a Schematic Phase Diagram

The schematic phase diagram shown in Fig. 1d was created from a simple model of the Gibbs free energy of the various phases present in the perovskite solution (DJ n=1-4, δ-FAPbI$_3$, α-FAPbI$_3$, and a solvated phase). Free energies were plotted together on a HI-DJ n1-FAPbI$_3$ ternary plot, and the corresponding ternary phase diagram was obtained from the convex hull of this free energy landscape. Crystalline phases were given Kronecker delta – like free energy functions positioned at their crystal stoichiometry, with the height of the delta tuned to reproduce the observed order of crystal growth from solution ($|G(DJ\ n2)| > |G(n3)| > |G(n4)| > |G(n1)| > |G(\alpha\text{-FAPbI}_3)|$). The free energy of these phases was taken to be temperature-invariant, while $|G(\delta\text{-FAPbI}_3)|$ was made to decrease from $|G(DJ\ n2)|$ to 0 with temperature. The free energy of the solution phase was taken to be a convex function with a maximum near the center of the ternary plot and zeros at the corners and along the DJ – FAPbI$_3$ edge. The magnitude of the solution-phase free energy was increased with temperature.

The ternary phase diagram constructed from the free energy of these phases was made to evolve with temperature. A horizontal line (a line of constant [HI]) across the ternary phase diagram was chosen to represent the solution concentration, and the intersection of each region of the phase diagram with this line was tracked as temperature was increased. In this way the binary phase diagram representing FA DJ at a specific concentration in HI solution was constructed. The process of creating the binary phase diagram from ternary phase diagrams is shown in Fig. S4.

## 4.3 1D X-ray diffraction measurements

The measurements were conducted using a Rigaku Smartlab II X-Ray diffractometer with Cu(Kα) radiation (λ = 1.5406 Å), running at 40 kV and 44 mA.

## 4.4 Single crystal structure measurements

Intensity data of a black plate single crystal of thick-layered perovskite containing 3AMP were collected at 293 K. A suitable single crystal with dimensions of ~0.1×0.1×0.02 mm3 was mounted on a MiTeGen loop with Paratone oil on a STOE StadiVari diffractometer equipped with an AXO Ag Kα micro-focus sealed Xray A-MiXS source (λ = 0.560834 Å), running at 65 kV and 0.68 mA, and a Dectris Pilatus3 R CdTe 300K Hybrid Photon Counting detector. Data reduction was performed with the CrysAlisPro software using a spherical absorption correction. The structure was solved with the ShelXT structure solution program using the Intrinsic Phasing solution method and by using Olex2 as the graphical interface. The model was refined with ShelXL using least squares minimization.

### 4.5 Absorbance measurements

The optical absorbance measurements were conducted using a broad-band light source (Thorlabs Solis-3C) focused onto the sample with a 50 μm beam size. The transmitted spectrum was collected by optical fiber and then sent to the spectrometer (Andor Kymera 328i) and CCD (Andor iDus 416). The measurement was conducted on thin KCSC crystals.

### 4.6 PL measurements

The photoluminescence (PL) spectroscopy of FA DJ 2D perovskites was measured based on a lab-built confocal microscopy system. The sample was photo-excited at 480nm using a supercontinuum pulsed laser (repetition rate 78MHz, temporal width ~50ps, NKT Photonics) spectral selected at 480nm. The laser was focused onto the sample through a 50x objective (0.42 NA) with ~ 1um beam size, yielding an excitation intensity of $3.6\times10^4$ mW/cm$^2$. The PL data was collected from 500 to 950 nm (1.3-2.4eV) by a spectrometer (Andor Kymera 328i) and a CCD camera (Andor iDus 416). For room temperature PL measurements, the sample was kept and measured at ambient condition. For temperature-dependent PL spectroscopy, the sample was kept under vacuum ($10^{-4}$ to $10^{-5}$ torr) in a closed-cycle cryostat (Advanced Research Systems) with sample temperature range from 6.5K to 300K using a temperature controller (Lakeshore). The excitation intensity was kept the same for all temperature ranges.

The temperature-dependent reflectance measurements were carried out in the same microscopy system, with the broad-band white light source (Thorlabs Solis-3C) focused on the sample guided by a beam splitter. The reflected spectra were normalized by reflectance of silver mirror on the sample plane.

### 4.7 1D and 2D NMR

The 3D a-FAPBI$_3$ and 2D RP and DJ materials were separately packed into 1.3 mm (outer diameter) zirconia rotors fitted with VESPEL caps without any further sample pretreatment. Solid-state MAS NMR experiments were conducted on a 21.1 T (Larmor frequencies of $^1$H and $^{207}$Pb were 900.2 MHz and 188.6 MHz, respectively) Bruker AVANCE-NEO spectrometer using a double resonance 1.3 mm H-X probehead. Unless specified, the MAS frequency was 50 kHz in all ssNMR experiments. The 1D $^1$H MAS NMR spectra were acquired by co-addition of 16 transients, where the relaxation delays were optimized to ensure the quantitate analysis of peak integrals. The spin-lattice relaxation time ($T_1$) values are determined from saturation recovery measurements and analyses. Echo-detected 1D $^{207}$Pb MAS NMR experiments were carried out with 40960 co-added transients, using 1 rotor-period echo delay (20 microseconds) with a repletion delay of 1.2 s, leading to an experimental time of 14 h each. All 2D $^1$H-$^1$H spin diffusion NMR experiments were acquired using three-pulse noesy-like sequence under fast MAS with 50 ms and 500 ms of mixing times. A rotor-synchronized increment of 20 microseconds was applied to detect 400 $t_1$ increments. The $^1$H experimental shift was calibrated with respect to neat TMS using adamantane as an external reference ($^1$H resonance, 1.82 ppm). All spectra were processed using Bruker Topspin 4.1 inbuilt package.

## 4.8 Low-frequency Raman measurement

Raman spectroscopy measurements were carried out using a diode laser module stabilized by volume holographic grating (VHG) filter, operating at 830nm (Coherent Ondax THz-Raman). The excitation laser was introduced into a home-built microscope and focused onto the sample using a microscope objective (Nikon ELWD 20x, NA 0.45). Typical laser power on sample surface was 2.4 mW. The light retro-reflected from the sample was routed back into the laser module which integrates a set of VHG notch filters. The Rayleigh scattered light was attenuated by the VHG filter set. The remaining Raman scattered light was subsequently fiber coupled into a spectrometer (Princeton Instruments Iso-Plane 320), dispersed by a 1200 grooves/mm grating, and captured by a thermoelectrically cooled CCD camera (PIXIS-BRX400). Typical integration time was 100s.

## 4.9 Computational methods

DFT calculations for obtaining the PDF, phonon spectral functions, and DOS were performed using the Quantum Espresso (QE) software[21,22]. A kinetic energy cutoff of 120 Ry, the PBEsol approximation[23], and optimized norm-conserving Vanderbilt pseudopotentials[24] were used. To obtain the polymorphous structures[10] we follow the method described in Ref. [[25]]. We employ 2x2x2 and 2x2x1 supercells of the primitive cells of cubic $FAPbI_3$ and $(3AMP)FAPb_2I_7$ containing 96 and 164 atoms, respectively. Geometry optimizations in supercells were performed by allowing the nuclear coordinates to relax and keeping the lattice constants fixed to their experimental values. We employed 3x3x3 and 3x3x2 uniform k-grids to sample the Brillouin Zone of polymorphous cubic $FAPbI_3$ and $(3AMP)FAPb_2I_7$. Geometry optimizations and phonon calculations (see below) were performed using scalar relativistic pseudopotentials, neglecting the effect of spin-orbit coupling.

*Doubly Screened Hybrid*: We perform density functional based calculation with the projector augmented wave potentials[26] as implemented in the VASP code[27]. The energy cut-off for the expansion of the wave-functions is set at 450 eV, and spin-orbit coupling interactions were taken into account for all calculations. To maintain the symmetry of the structures, we proceed to replace the organic moieties with Cs atoms that were placed at the position of the N-atoms. We checked that this replacement is not affecting the electronic structure by comparing the band structures, and that no symmetry is broken (i.e., the space group remains the same as the experimentally observed). Furthermore, to include the effect of polymorphism is the structures, we also performed a calculation of a 2x2x2 super-cell of $FAPbI_3$ (poly-$FAPbI_3$) which include the FA molecules. To overcome the well-known band-gap underestimation of DFT, we employ the doubly screened hybrid exchange-correlation functional.[28–30] We calculate from first-principles the $\varepsilon_\infty$ (details below) for all compounds and use $\varepsilon^{-1}_\infty$ as the mixing parameter for the long range, while the exact exchange is used for the short range. For the employed model dielectric function, we used a screening length parameter μ value of 1.05, which was taken from $GW_o$ calculations of Bokdam et al. on $FAPbI_3$.[31] For the calculations of the exact exchange we employed uniform Γ-centered k-point grids of 4x4x2 for the layered materials, and 6x6x6 (2x2x2) for the mono-$FAPbI_3$ and poly-$FAPbI_3$, respectively.

*Calculation of $\varepsilon_\infty$*: We used the density functional perturbation theory (DFPT) to calculate the dielectric constant of the n=1,2,3,4, mono-$FAPbI_3$ and poly- $FAPbI_3$ compounds. We employed Γ-centered k-point grids of 12x12x4, 12x12x4, 10x10x4,10x10x4, 20x20x20 and 6x6x6, respectively. Local field effects were included at the DFT level. For the layered materials we replaced the organic spacers with N atoms, thus to avoid artifacts due to the different dielectric screening, we used a capacitor stack model to extract the $\varepsilon_\infty$ of the inorganic slab layer from the DFPT calculation.[28,32] The final $\varepsilon_\infty$ for the materials are given in Table S4.

## 4.10 Phonons and PDF

The phonons of the polymorphous structures were calculated by means of finite differences using the zeroth order iteration of the anharmonic special displacement method (A-SDM)[33]. PDFs of the polymorphous structures were evaluated as

$$G(\tau) = \frac{1}{\tau}\sum\sum \delta\left(\tau - \tau_{\kappa\kappa\prime}\right)$$

by replacing the delta function with a Gaussian of width 0.035 Angstroms. Here, $\kappa$ is the atom index and $\tau_{\kappa\kappa\prime}$ defines the distance between atom κ and κ'. To include the effect of thermal vibrations in the PDFs we employed thermally displaced configurations in 10x10x10 and 20x20x1 supercells. The configurations were generated using the A-SDM at essentially no cost, as it takes advantage of Fourier interpolation of the phonons in the reciprocal space.[34]

## 4.11 Phonon spectral functions

Phonon spectral functions were calculated using the phonon unfolding technique[25] as implemented in the ZG package of EPW.[35] We employed 663 and 787 equally-spaced **q**-points for (3AMP)FAPb$_2$I$_7$ and FAPbI$_3$, respectively, and $12 \times 12 \times 12$ **g**-grid of reciprocal lattice vectors to ensure convergence of the spectral weights.

## 4.12 DOS at finite temperatures

Electronic structure calculations for the DOS were performed using fully relativistic pseudopotentials, including the effect of spin-orbit coupling. To calculate the DOS at the VBM and CBM for 300 K, we employed the A-SDM. In the A-SDM, anharmonic phonons computed for the polymorphous structures are used to generate thermally displaced configurations, which capture the effect of electron-phonon coupling on the electronic structure when combined with DFT calculations.[25] To obtain the DOS of polymorphous FAPbI$_3$ and (3AMP)FAPb$_2$I$_7$ at 300 K we used 10 configurations of supercell size 4x4x4 (768 atoms) and 4x4x1 (656 atoms), respectively. We used 1x1x1 and 1x1x2 grids to sample the Brillouin zone of these supercells. The phonon-induced band gap renormalization, $\Delta\varepsilon_g$, with respect to the band gap of the polymorphous structure at static-equilibrium was determined by taking the average renormalization over the 10 configurations.

## 4.13 PL diffusion measurements

Montana Instruments Cryostation, with attached turbo pump, was employed to maintain a high vacuum environment ($10^{-4}$ to $10^{-5}$ Torr) during room temperature (295 K) measurements. An objective with 100× magnification and 0.9 NA focused a 532 nm continuous-wave laser onto the samples with a resulting beam diameter (D4σ) of ~ 0.6 μm. A neutral density filter was placed in the beam path, before the objective, to modulate the laser power for the power-dependent measurements. Laser contributions from the reflected PL were blocked with a 550 nm long-pass filter, installed before the detector. Additionally, free carrier diffusion contributions were removed with a 700 nm short-pass filter for both n=2 and n=3 FA DJ perovskites. Subsequent real-space emission images of the sample surface were acquired using an EMCCD camera in low electron multiplication mode[36]. An n-type GaAs wafer was used for calibration of measurements and to confirm reliability of analysis.

### 4.14 Time-resolved Photoluminescence

The TRPL spectrum of DJ n=3 sample was acquired with a lab-built confocal microscopy system. The sample was photo-excited using a 50-ps-pulsed super-continuum laser (NKT Photonics, repetition rate tuned at 39MHz) spectral selected at 480 nm. The excitation laser was focused onto the sample with ~ 1um beam size and average excitation intensity of $1.8 \times 10^4$ mW/cm$^2$. The emission was collected using PicoQuant HydraHarp 400 time-correlated single photon counting system combined with an Avalanche Photo-Diode (MPD-SPAD). The emission was spectrally filtered through a spectrograph (Andor Kymera 328i) to remove laser excitation. During TRPL measurement, the sample was kept under vacuum ($10^{-4}$ to $10^{-5}$ torr) in a closed-cycle cryostat and maintained at cryogenic temperature (T = 6.5K).